\documentclass[twocolumn]{aastex701}
\graphicspath{{./}{}{figures/}}
\usepackage[T1]{fontenc}
\usepackage{amsmath}

\begin{document}

\title{Using Structure Functions to Characterize Protoplanetary Disk Velocity Residuals}

\author[0000-0003-1534-5186]{Richard Teague} 
\affiliation{Department of Earth, Atmospheric and Planetary Sciences, Massachusetts Institute of Technology, Cambridge, MA 02139, USA}
\email[]{rteague@mit.edu}

\author[0009-0000-7872-3493]{Sean M. Andrews}
\affiliation{Center for Astrophysics | Harvard \& Smithsonian, Cambridge, MA 02138, USA}
\email[]{}

\author[0000-0001-7258-770X]{Jaehan Bae}
\affiliation{Department of Astronomy, University of Florida, Gainesville, FL 32611, USA}
\email[]{}

\author[0000-0001-6378-7873]{Marcelo Barraza-Alfaro}
\affiliation{Department of Earth, Atmospheric and Planetary Sciences, Massachusetts Institute of Technology, Cambridge, MA 02139, USA}
\affiliation{Niels Bohr International Academy, Niels Bohr Institute, Blegdamsvej 17, DK-2100 Copenhagen Ø, Denmark}
\email[]{}

\author[0000-0003-1333-2840]{Abigail Bodner} 
\affiliation{Department of Earth, Atmospheric and Planetary Sciences, Massachusetts Institute of Technology, Cambridge, MA 02139, USA}
\email[]{}

\author[0000-0001-9656-7682]{Jonathan D. Henshaw}
\affiliation{Max Planck Institute for Astronomy, K\"onigstuhl 17, D-69117 Heidelberg, Germany}
\email[]{}

\author[0000-0003-4853-5736]{Giovanni Rosotti} 
\affiliation{Dipartimento di Fisica, Universit\`a degli Studi di Milano, Via Celoria 16, 20133 Milano, Italy}
\email[]{}

\begin{abstract}
Protoplanetary disks, the birthplace of planetary systems, are highly dynamic environments. High angular and spectral resolution observations have revealed non-Keplerian motions in nearly every source studied, indicative of (magneto-)hydrodynamical instabilities or planet-disk interactions. However, these velocity perturbations rarely exhibit ordered structure, limiting our ability to attribute them to a specific mechanism. We propose the second-order structure function as a tool for recovering structural properties of velocity perturbations in the presence of significant noise. Using morphological models, we demonstrate how easy-to-calculate heuristics and parametric model fits together identify structure in noisy data and recover its properties, including correlation lengths, their radial scaling where genuinely present, and the anisotropy. We further outline a workflow to de-noise the data and to build a matched null model against which inferences are compared. Applying the framework to one numerical simulation each of the gravitational, magneto-rotational and vertical-shear instabilities, we recover their structure at high significance and the structural heuristics separate all three pairs, with the gravitational and vertical-shear instability simulations remaining the hardest to distinguish once observationally degraded. As a demonstration on real data, $^{12}$CO observations of TW~Hya reveal azimuthally elongated velocity structure that persists once the large-scale $m=1$ component is removed. We also delineate where a Gaussian description breaks down for coherent features such as spiral arms, and how inclination affects the recovered heuristics. The methodology is implemented in the \texttt{eddy} package and provides a foundation for statistically characterizing turbulent velocity structure in high-resolution disk observations.
\end{abstract}

\section{Introduction} 

It is now clear that the gas dynamics in protoplanetary disks exhibit substantial deviations from a simple Keplerian-like velocity profile \citep[e.g.,][]{Pinte_ea_2023}. While for some sources the perturbations form a clear structure, easily attributable to known (magneto-)hydrodynamical processes, such as the multi-armed spirals indicative of the gravitational instability in AB~Aur \citep{Speedie_ea_2024}, for many sources the structure is far less interpretable \citep[e.g.,][]{Fukagawa_ea_2026}. This is particularly true of the velocity signatures expected from (magneto-)hydrodynamical turbulence, which are predicted to be stochastic and spatially correlated rather than to form large-scale, coherent, easily parameterized morphology \citep[e.g.,][]{Heinemann_Papaloizou_2009, Barraza-Alfaro_ea_2025}. Characterizing such fields calls for statistical descriptors of the correlation structure of the residuals rather than the fitting of individual features.

In this paper we explore the use of the second-order structure function, $S_2$, as a way to identify and characterize structure in an apparently noise-dominated velocity map. First introduced in Kolmogorov's description of the inertial range of incompressible turbulence \citep{Kolmogorov_1941}, the structure function is a standard descriptor of turbulent and geophysical flows and is widely used in terrestrial applications such as the study of atmospheric and oceanographic turbulence \citep{Monin_Yaglom_1975}. It has likewise seen extensive use in astronomical contexts, most prominently in characterizing turbulence in the interstellar medium and molecular clouds, including the size--linewidth relation of \citet{Larson_1981}, the velocity structure of molecular gas \citep{Ossenkopf_MacLow_2002, Heyer_Brunt_2004} as well as in recovering the underlying velocity statistics from line-of-sight velocity centroids \citep{Esquivel_Lazarian_2005} and in quantifying fluctuations in the solar wind and magnetized astrophysical plasmas \citep{Bruno_Carbone_2013}.

Given the long-standing assumption that protoplanetary disks exhibit some level of turbulence \citep{Lesur_ea_2023}, our goal is to establish what the structure function can and cannot reliably recover from a disk velocity field, before applying it to real data. To this end we work with synthetic, correlated Gaussian random fields drawn natively on the disk polar grid, for which the true correlation structure is known by construction. We introduce a set of empirical heuristics together with a complementary parametric fit, and we build a noise-aware pipeline that de-noises at the level of $S_2$ and assesses the significance of each measured quantity against matched noise-only null ensembles. With this machinery, implemented in the \texttt{eddy} package \citep{eddy}, we quantify the biases intrinsic to the method and determine which properties of the field remain recoverable in the presence of noise.

The layout of this manuscript is as follows. We start in Section~\ref{sec:morphological_models} by describing the models we will use for the demonstration of this methodology. In Section~\ref{sec:structure_function} we introduce the structure function and build intuition through application to simple models. These models are extended in Section~\ref{sec:noisy_models} where we consider the impact of finite angular resolution and noise in our characterization. In Section~\ref{sec:numerical_simulations} we apply the complete pipeline to velocity fields from numerical simulations of (magneto-)hydrodynamical instabilities, testing whether the velocity fields of these simulations can be identified and distinguished from one another. In Section~\ref{sec:twhya} we illustrate what the method recovers from real data, using observations of TW~Hya, and in Section~\ref{sec:inclined} we discuss how it extends to inclined disks. We summarize our findings in Section~\ref{sec:conclusions}.

\section{Morphological Models}
\label{sec:morphological_models}

When faced with velocity residuals that are not dominated by well defined morphologies, we must resort to statistical descriptors of the field to characterize the (potential) structure within it. With a view to building our intuition for such analyses, we start by introducing a family of morphological models that are able to reproduce the main features of known \hbox{(magneto-)}hydrodynamical instabilities \citep[e.g.,][]{Barraza-Alfaro_ea_2025}. Such models allow us to quickly and efficiently draw large ensembles of realizations in order to stress test the proposed methodology. Importantly, these models do not consider the effects of projection along the line of sight which lead to periodic azimuthal variations in strength \citep[e.g.,][]{Teague_ea_2022} and therefore should be considered as simple models for demonstration's, rather than interpretation's, sake.

\subsection{Gaussian Random Fields}
\label{sec:morphological_models:GRFs}

We use correlated, zero-mean Gaussian random fields (GRFs) which we draw on the polar $(r,\, \phi)$ grid. Correlations between two points at similar radii (such that the small angle approximation holds), separated in polar coordinates by a distance $(\delta r,\, \delta\phi)$ are described by the correlation kernel
\begin{equation}
  C(\delta r,\delta\phi) = \sigma^2 \exp\!\left( -\frac{\delta r^2}{2\ell_r^2} -\frac{(r\,\delta\phi)^2}{2\ell_\phi^{2}} \right),
  \label{eq:kernel}
\end{equation}
where the azimuthal term uses the physical arc length $r\,\delta\phi$. We allow the two correlation lengths to scale with radius as
\begin{equation}
  \ell_r(r) = \ell_{0,r}\,(r/r_0)^{\alpha_r},
  \label{eq:radial_scale_length}
\end{equation}
and
\begin{equation}
  \ell_\phi(r) = \ell_{0,\phi}\,(r/r_0)^{\alpha_\phi}.
  \label{eq:azimuthal_scale_length}
\end{equation}
Note that the kernel scale here, $\ell$, is related to the FWHM of a convolution kernel via $\ell = \theta_b / 2\sqrt{\ln 2} \approx 0.60\, \theta_b$. To generate a model, realizations of $g \sim \mathcal{N}(0, C)$ are drawn using the non-stationary covariance of \citet{Paciorek_Schervish_2006}, and validated against the analytic structure function predicted by the kernel parameters.

\begin{deluxetable}{lcccccc}
    \tablecaption{Parameters of the four morphological models. \label{tab:morphological_models}}
    \tablewidth{\columnwidth}
    \tablehead{
        \colhead{Model} &
        \colhead{$\sigma$} &
        \colhead{$\ell_{0,r}$} &
        \colhead{$\ell_{0,\phi}$} &
        \colhead{$\alpha_r$} &
        \colhead{$\alpha_\phi$} &
        \colhead{$\beta$} \\
        \colhead{} &
        \colhead{} &
        \colhead{(\arcsec)} &
        \colhead{(\arcsec)} &
        \colhead{} &
        \colhead{} &
        \colhead{(\degr)}
    }
    \startdata
    A (isotropic, stationary) & 1 & 0.06 & 0.06 & 0.0 & 0.0 & 0 \\
    B (isotropic, power-law) & 1 & 0.06 & 0.06 & 1.0 & 1.0 & 0 \\
    C (anisotropic) & 1 & 0.06 & 0.30 & 0.0 & 0.0 & 0 \\
    D (anisotropic + spirals) & 1 & 0.06 & 0.30 & 0.0 & 0.0 & 10 \\
    \enddata
    \tablecomments{All four models are normalized to a reference radius of $r_0 = 1\arcsec$ in Equations~\ref{eq:radial_scale_length} and \ref{eq:azimuthal_scale_length}. Correlation lengths are the kernel $\ell$ of Equation~\ref{eq:kernel}; the equivalent generating beams have FWHM of $0\farcs1$ and $0\farcs5$, respectively.}
    \end{deluxetable}

We recover an isotropic field when $\ell_{0,r} = \ell_{0,\phi}$ and $\alpha_r = \alpha_\phi$, and a stationary field when $\alpha_r = \alpha_\phi = 0$ such that the scale lengths are independent of $r$. In addition to the stationarity and anisotropy of the field, we also allow for some flocculent spiral structure through a tilting of the correlation kernel by a given pitch angle, $\beta$:
\begin{equation}
    C(\boldsymbol{\delta}) = \sigma^2 \exp\! \left( -\frac{1}{2}\boldsymbol{\delta}^{\top} \mathbf{R}(\beta)\, \boldsymbol{\Lambda}^{-1}\mathbf{R}(\beta)^{\top} \boldsymbol{\delta} \right),
\end{equation}
where $\boldsymbol{\delta} = (\delta r,\, r\delta \phi)^{\top}$, $\mathbf{R}(\beta)$ is the rotation relative to $\hat\phi$, and $\boldsymbol{\Lambda} = {\rm diag}(\ell_\phi^2,\, \ell_r^2)$. The elongated correlation patches then trail at angle $\beta$ from the azimuthal direction, producing flocculent spiral structure with random arm phases.

\subsection{Model Suite}
\label{sec:morphological_models:models}

\begin{figure*}
    \includegraphics[width=\textwidth]{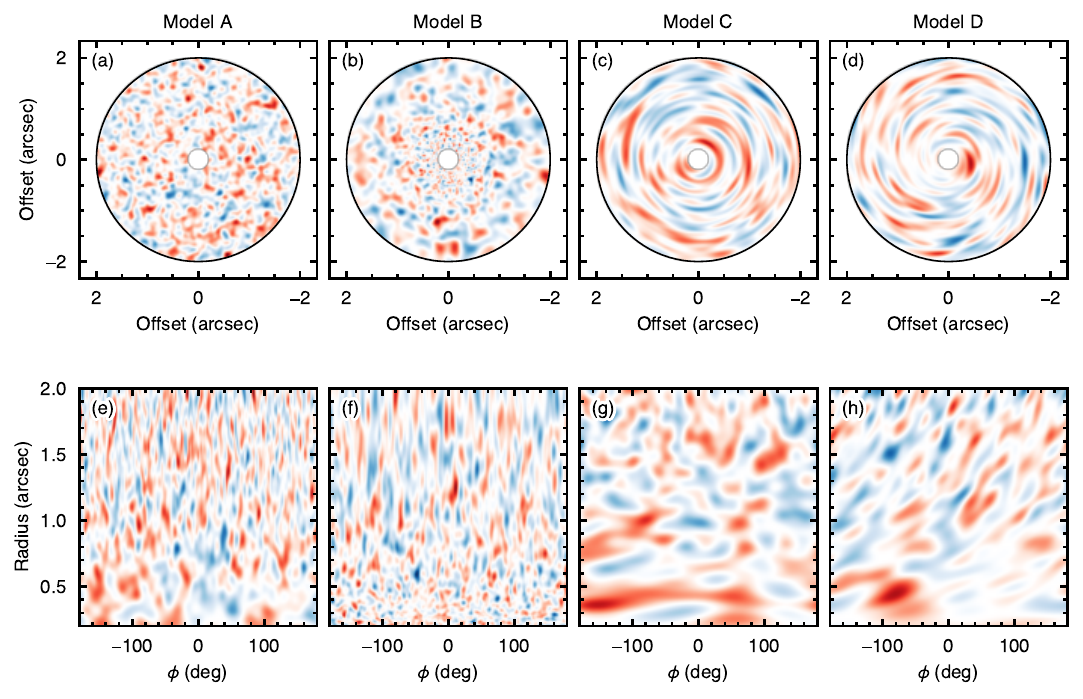}
    \caption{On-sky, top row, and polar projections, bottom row, of the four phenomenological models used to demonstrate the structure function. The colors represent velocity perturbations normalized such that each model has a zero mean and a standard deviation of unity. Note how the polar projection stretches structure out azimuthally at small radii. The on-sky projections assume an inclination of $i = 0\degr$.}
    \label{fig:structure_models}
\end{figure*}

For this experiment we consider four families of models, Model A through D, which are designed to coarsely mimic structures expected from numerical simulations and are summarized in Table~\ref{tab:morphological_models}. Model A represents correlated white noise and serves as a `null' model. In this model, the correlation lengths are set to $\ell_{0,r} = \ell_{0,\phi} = 0\farcs060$, independent of radius. Model B takes the same scale normalization, but grows this isotropically with radius through $\alpha_r = \alpha_\phi = 1.0$. A radial scaling of the perturbations is expected in reality, as the dominant eddy-scales should track the local scale height of the disk which also grows with radius in the disk. Model C implements strong azimuthal correlations that produce annular features reminiscent of the vertical shear instability \citep[although without the radially varying corrugation length;][]{Nelson_ea_2013}. This assumes a stationary field, but adopts an azimuthal correlation length five times that of the radial correlation length: $\ell_{0,\phi} = 5\,\ell_{0,r} = 0\farcs3$. Finally, Model D introduces spirals with a non-zero pitch angle, $\beta = 10\degr$, while adopting the same parameters as Model C. A single realization of each is displayed in Figure~\ref{fig:structure_models} with both on-sky (assuming a completely face-on orientation) and polar projections.

An important aspect of this modeling framework is that multiple components can be combined to generate an arbitrarily complex field. As described in the following section, appropriate statistical techniques allow us to decompose these multiple components.

\section{Structure Functions}
\label{sec:structure_function}

The second order structure function describes the average variance in a field at separation $\ell$, given by
\begin{equation}
    S_2(\ell) = \langle |f(\mathbf{x}) - f(\mathbf{x} + \ell)|^2 \rangle,
\end{equation}
where $f$ is an arbitrary scalar field and the angle brackets denote an average over all pairs of points separated by $\ell$. This can be considered a spatial analog to a power spectrum where the scales are real spatial separations rather than wavelengths. For the case of protoplanetary disks it is useful to consider orthogonal lags, $\delta r$ and $\delta \phi$, as the subsequent interpretation generally considers radial and azimuthal variations separately.

One subtlety in the application to protoplanetary disks is that we expect the underlying field to be non-stationary: the scale of perturbations should be pegged to the local scale height of the disk, and thus vary (most likely increasing) with radius. In the parlance of our GRF model, $\alpha \neq 0$. To account for this, we consider two forms of the structure function: a \emph{global} $S_2$, which pairs points across the whole disk and so mixes correlation scales from every radius together, and a \emph{stack} of $S_2$, in which each element instead restricts pairs to a narrow reference annulus of half-width $\Delta r$ centered on a given radius $r_0$, under the assumption that the field is locally, if not globally, stationary. Repeating this at a sequence of reference radii lets us track how the correlation lengths evolve with $r$ allowing us to test stationarity. We adopt these terms to align with the naming convention used in \texttt{eddy} \citep{eddy}, which implements these as \texttt{StructureFunction} and \texttt{StructureFunctionStack}, respectively.

\subsection{Building Intuition}
\label{sec:structure_function:intuition}

\begin{figure*}
    \centering
    \includegraphics[width=0.9\textwidth]{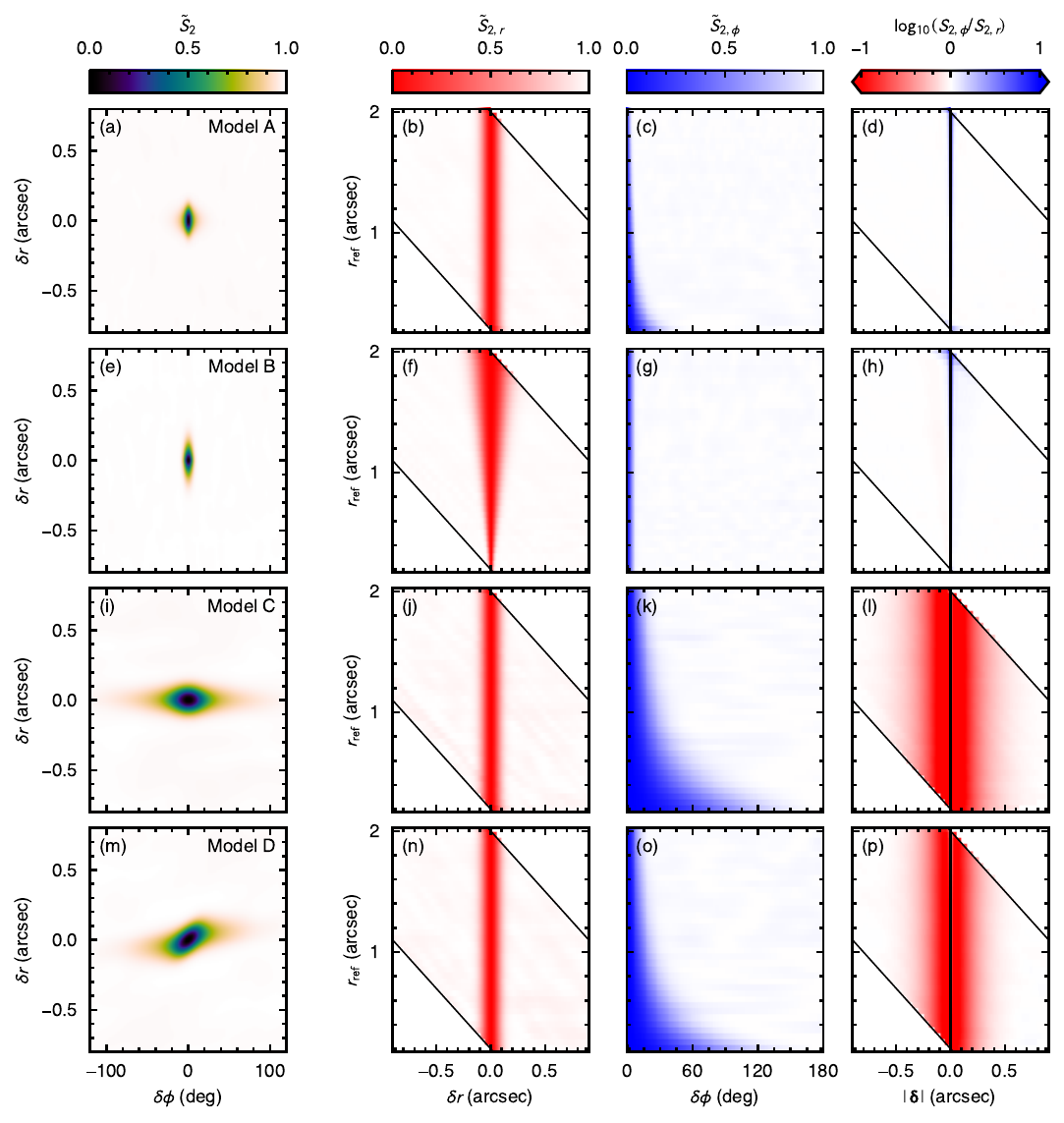}
    \caption{An atlas of normalized structure functions for the four different phenomenological models. All panels show the mean of 1{,}000 draws in order to suppress noise from sampling variance. The left column shows the normalized global $\tilde{S}_2$, while the three remaining columns show the row-normalized radial structure function, $\tilde{S}_{2,r}$, row-normalized azimuthal structure function, $\tilde{S}_{2,\phi}$, and the structure-function ratio, $S_{2,\phi}\, / \,S_{2,r}$, for the four morphological models. In these figures, brighter colors represent a smaller $S_2$ and thus a higher level of correlation and whiter regions show uncorrelated regions.}
    \label{fig:s2_atlas}
\end{figure*}

To build intuition for how $S_2$ encodes properties of the velocity field, we plot the global $S_2$, the row-normalized radial structure function $\tilde{S}_{2,r}$ (where the tilde denotes row-normalization, useful for visualization purposes), the row-normalized azimuthal structure function, $\tilde{S}_{2,\phi}$, and the ratio of the two, $S_{2,\phi} \, / \, S_{2,r}$, for the four models in Figure~\ref{fig:s2_atlas}. Note that for the calculation of this ratio the azimuthal structure function is converted from angular lags to arc-length lags to match the radial structure function. We plot the average of $N = 1{,}000$ draws of the GRF in order to suppress the impact of sampling variance. A discussion of how to contend with this in the presence of noise is postponed to Section~\ref{sec:noisy_models}.

The first column shows the global $S_2$ for each model. A low $S_2$ value means there is little variance between points separated by that specific lag. We see that $S_2$ approaches 0 as $\ell \rightarrow 0$; adjacent points are highly correlated. At large lags, the variance saturates to the overall variance of the field as the points are no longer correlated. In the absence of long-range correlations, this plateau sits at $2\sigma^2$. For Models A, B and C, $S_2$ is aligned along the two axes, while for Model D, the rotation introduced by the non-zero $\beta$ manifests as a twisting of the core. Importantly, spiral information is encoded in the off-diagonal components of the global $S_2$ and is washed out in other projections.

The second and third columns show the row-normalized radial and azimuthal structure functions as a function of reference radius. This projection is useful to identify the stationarity of the field: if the field is stationary then the structure function should be independent of the reference radius and direction of lag (i.e., lags looking radially outwards versus radially inwards). This can be most clearly seen when comparing $\tilde{S}_{2,r}$ for Model A and B, shown in panels (b) and (f), where Model B shows a growing correlation length with reference radius (Model B is the only model with $\alpha_r \neq 0$). The azimuthal structure function is more complex. Although most models are drawn assuming a stationary field in the azimuthal direction ($\alpha_\phi = 0$), the polar-coordinate representation of the field results in a stretching of the azimuthal structure at small radii, and a compression at larger radii. This introduces a level of non-stationarity that manifests as large correlation lengths at small reference radii which shrink with radius, i.e., an effective negative $\alpha_\phi$. Interestingly, Model B shows a close to stationary field for $S_{2,\phi}$ as the intrinsic azimuthal correlation length scales with radius, negating the polar-coordinate representation bias.

We show this ratio, a measure of the anisotropy of the field, in the fourth column. Plotted as $\log_{10}(S_{2,\phi} / S_{2,r})$, a zero value denotes a purely isotropic field, a negative value means that $S_{2,r} > S_{2,\phi}$ and so there is more extended structure in the azimuthal direction than radial, such as in Models C and D, and a positive value means structures are primarily in the radial direction. Only Models C and D were generated with anisotropies ($\ell_{0,r} \neq \ell_{0,\phi}$), and that is readily apparent from the $\log_{10}(S_{2,\phi} / S_{2,r})$ maps. Anisotropy is only found at small lags, as at large lags the field becomes uncorrelated and thus there is no preferred scale. At small lags, where both structure functions grow as the square of the lag, the ratio tends to $\mathcal{A}^{-2}$, where $\mathcal{A} = \ell_\phi / \ell_r$ is the anisotropy defined in Section~\ref{sec:structure_function:heuristics}.

\subsection{Empirical Characterization}
\label{sec:structure_function:heuristics}

Although examining plots of the structure builds intuition, a set of quick-to-calculate heuristics allows us to summarize the information in these panels into readily-comparable values. We propose a set of six heuristics, motivated by the model parameters described in Section~\ref{sec:morphological_models:GRFs}, that allow us to probe properties such as magnitude and scale of structure in the field, its stationarity and isotropy: $\{\hat{\sigma},\, \hat{\ell}_r,\, \hat{\ell}_s,\, \hat{\alpha}_r,\, \hat{\alpha}_\phi,\, \hat{\mathcal{A}}\}$, where we include a hat to differentiate these heuristics from the model input parameters. These can be quickly calculated within \texttt{eddy} using the newly implemented \texttt{measure\_heuristics} method.

\subsubsection{Magnitude and Scale}

The most basic question we can ask of our data is whether there is structure that is inconsistent with our noise model. We can derive the average amplitude of our variations, $\hat{\sigma}$, by rescaling the radius-averaged $S_2$ plateau which should be equal to $2\sigma^2$. We can also derive the typical correlation length in the radial and azimuthal direction, $\hat{\ell}_r$ and $\hat{\ell}_s$, by taking the radially averaged half-power lag (the lag at which $S_2$ reaches half of its plateau value). Note that $\ell_s$ is the arc length, not the angular correlation length. When taking the radial average, each annulus is weighted by $n_{\rm eff} \approx 360\degr \cdot (r_{\rm out} - r_{\rm in}) \, / \, \ell_r \ell_\phi$, its effective number of independent correlation patches, which approximates inverse-variance weighting — not the raw pair count, which would over-trust annuli whose many pairs are correlated. These three heuristics can then be readily compared to the noise model to determine whether there is structure in the data. For example, the noise in a signal-free channel can have its RMS measured and compared to $\hat\sigma$, and the beam compared to $\hat{\ell}_r$ and $\hat{\ell}_s$ once converted. The appropriate conversion depends on how the noise arises. For a field made by convolving white noise with a beam of FWHM $\theta_b$, as we do throughout this work, the correlation length is that of the beam's autocorrelation, $\ell = \theta_b / 2\sqrt{\ln 2}$. For thermal noise in an interferometric image the autocorrelation is instead the point spread function itself, giving $\ell = \theta_b / 2\sqrt{2 \ln 2}$, shorter by a factor of $\sqrt{2}$. The second is the relevant one when comparing against real data.

\subsubsection{Stationarity}

\begin{figure*}
    \centering
    \includegraphics[width=\textwidth]{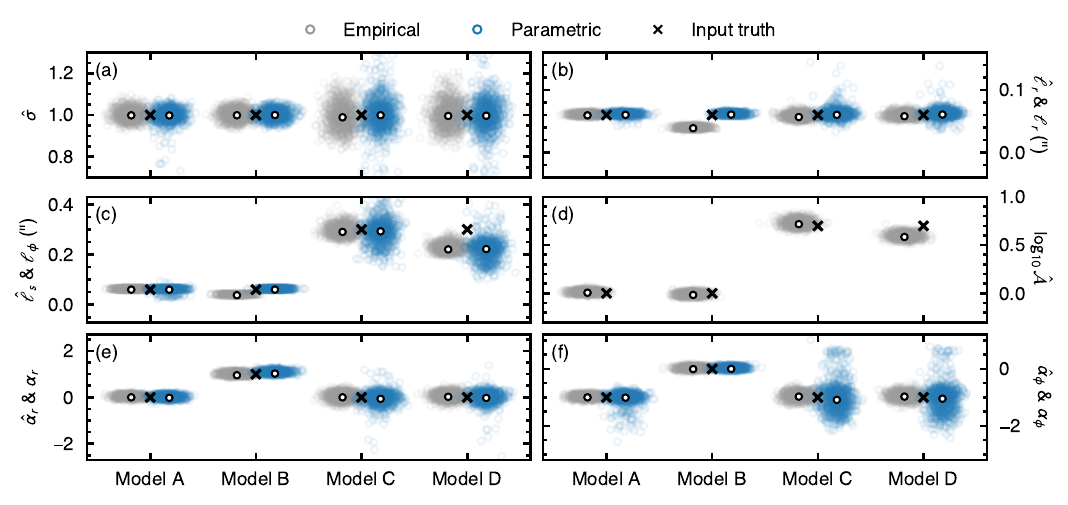}
    \caption{The ability to recover the properties of different GRF models using both empirical (gray) and model-fitting (blue) approaches. Each panel represents a different model property, while each column corresponds to a different model described in Section~\ref{sec:morphological_models:models}. The outlined circles in each point cloud mark the sample median. The black crosses show the true input value. As noted in the text, $\hat{\mathcal{A}}$ is not included in the parametric model. Panel (f) is drawn on the heuristic convention of Section~\ref{sec:structure_function:heuristics}, for which physical stationarity is $\hat\alpha_\phi = -1$; the fitted $\alpha_\phi$ and the input value of Table~\ref{tab:morphological_models} are both shifted by $-1$ so that the two estimators share the axis.}
    \label{fig:model_heuristics_vs_grf}
\end{figure*}

The indices $\hat\alpha_r$ and $\hat\alpha_\phi$ encode the stationarity of the field, capturing how the radial and azimuthal correlation lengths vary with reference radius. $\hat\alpha_r$ is the $n_{\rm eff}$-weighted slope of a linear fit of $\ln\hat\ell_r$ against $\ln r$. $\hat\alpha_\phi$ is measured the same way, but on the underlying \emph{angular} half-power width $\Delta\hat\phi_{1/2}$ (degrees) rather than on the reported arc-length heuristic $\hat\ell_s$ itself. Since $\ell_s = r\,\Delta\phi_{1/2}$, the slope of $\ln\Delta\hat\phi_{1/2}$ against $\ln r$ is offset by one power of $r$ relative to the scaling of the arc length.

A radius-independent radial scale gives $\hat\alpha_r = 0$. For the azimuthal direction the equivalent physical stationarity instead gives $\hat\alpha_\phi = -1$, since a constant arc subtends a shrinking angle with radius; $\hat\alpha_\phi = 0$ describes a self-similar field whose angular extent is constant.

\subsubsection{Anisotropy}

We define the anisotropy of the field as the ratio of the two correlation lengths, $\mathcal{A} = \ell_\phi / \ell_r$, so that $\mathcal{A} > 1$ describes structure elongated azimuthally, $\mathcal{A} < 1$ structure elongated radially, and $\mathcal{A} = 1$ an isotropic field. Because a longer correlation length along a direction keeps pairs correlated and so \emph{lowers} $S_2$ along it, this runs in the opposite sense to the structure-function ratio $S_{2,\phi} / S_{2,r}$ of Section~\ref{sec:structure_function:intuition}. Because the azimuthal correlation length is a physical arc length, $\ell_\phi = r\,\Delta\phi$, the measured angular scale must be converted to a length using the radius of each annulus before the ratio is formed. We therefore evaluate $\ell_\phi / \ell_r$ annulus by annulus where this conversion is exact, and take its $n_{\rm eff}$-weighted geometric mean, $\hat{\mathcal{A}} = \exp \langle \ln(\ell_\phi / \ell_r) \rangle$, rather than dividing the separately averaged $\hat{\ell}_r$ and $\hat{\ell}_s$. Averaging in $\ln\mathcal{A}$ both respects the multiplicative nature of the ratio and makes the estimator symmetric about the isotropic case.

As constructed, $\hat{\mathcal{A}}$ is the typical anisotropy averaged over the radial range considered; any radial \emph{variation} of the anisotropy is not captured by $\hat{\mathcal{A}}$ itself but is instead encoded in the radial scaling indices $\hat\alpha_r$ and $\hat\alpha_\phi$. Since $\hat{\mathcal{A}} = \ell_\phi / \ell_r \propto r^{\,\hat\alpha_\phi - \hat\alpha_r + 1}$, a radius-independent anisotropy corresponds to $\hat\alpha_r - \hat\alpha_\phi = 1$, with departures from this tracing how the elongation changes with radius. Because the arc-length conversion folds in the ring radius, $\hat{\mathcal{A}}$ also shares the polar-representation sensitivity of $\hat\alpha_\phi$.

\subsection{Parametric Characterization}
\label{sec:structure_function:parameterization}

In addition to purely empirically derived quantities, we can attempt to model aspects of the observed structure function under the assumption that the GRF parameterization described in Section~\ref{sec:morphological_models:GRFs} is a reasonable representation of the data. One particular advantage of the modeling approach over the purely empirical heuristics is that it is possible to include the rotation $\beta$ in the model, something that is lost when considering only the on-axis components (Model C and D are effectively indistinguishable in this space).

We use \texttt{eddy} to fit the observed $S_2$ with a correlation kernel described in Equation~\ref{eq:kernel} through the \texttt{fit\_GRF} method. This uses a least-squares optimization (\texttt{scipy.optimize.least\_squares}) with two iterations where the second iteration uses the best-fit model to define the weights (an iteratively reweighted least squares approach). This two-step process is necessary as we found empirical weighting resulted in an underestimated noise level, particularly at larger lags, resulting in a negative bias in other recovered parameters. The returned uncertainties are only the formal statistical uncertainties and substantially underestimate the true uncertainty. As the anisotropy is a derived model property, we do not plot it in Figure~\ref{fig:model_heuristics_vs_grf}. The characteristic correlation lengths are also re-scaled to be at $1\arcsec$ to ease comparison, however the fitting is performed with $r_0$ being the logarithmic center of the radial grid.

It is worth being explicit about when the fitted $\ell_{0,\phi}$ and the empirical $\hat{\ell}_s$ measure the same thing. Both are quoted in the kernel convention of Equation~\ref{eq:kernel}: the half-power lag is not reported raw, but is divided by $\sqrt{2\ln 2}$ inside \texttt{calculate\_heuristics} so that $\hat{\ell}_r$ and $\hat{\ell}_s$ carry the same $\ell$ that \texttt{fit\_GRF} returns, and therefore the same $\ell$ that relates to the FWHM of the generating beam through $\ell = \theta_b / 2\sqrt{\ln 2}$ (Section~\ref{sec:morphological_models:GRFs}). While they are directly comparable in principle, they are not the same estimator. $\hat{\ell}_s$ is the $n_{\rm eff}$-weighted average of $\ell_\phi(r)$ over the annuli entering the stack, whereas $\ell_{0,\phi}$ is the model value at the single reference radius $r_0$. The two therefore agree for a stationary field or when $r_0$ is placed at the effective radius of the fitted range. They differ significantly when the Gaussian kernel is misspecified, since $\hat{\ell}_s$ remains a well-defined descriptive lag whatever the shape of $S_2$, while $\ell_{0,\phi}$ inherits the assumption that a Gaussian envelope describes it.

\subsection{Application to Models}
\label{sec:structure_function:models}

To demonstrate the use of these empirical and parametric characterizations, we calculate them for each of the 1{,}000 realizations used to populate Figure~\ref{fig:s2_atlas} and compare them to their expected value in Figure~\ref{fig:model_heuristics_vs_grf}. We do not consider the impact of noise on these recoveries, instead deferring a discussion of this to Section~\ref{sec:noisy_models}. In Figure~\ref{fig:model_heuristics_vs_grf}, the gray circles show the empirical parameters for individual realizations, the blue circles show the $S_2$ model fit and the overlaid black-outlined circles show the sample median. The black crosses show the input or expected value.

One immediate point to notice is that there is a spread in derived quantities arising due to sampling variance. In a single realization, there are only a (small) number of independent pairs of points with which to measure $S_2(\boldsymbol{\delta})$, meaning that small number statistics can result in biases in the recovered value. This spread can be considered as an uncertainty on the measurement, similar to `cosmic variance' in cosmology. The two estimators are comparably precise in the amplitude, while they differ in precision for the correlation lengths, where the GRF fit suffers more from the sampling variance.

We find differences in the recovered values of correlation lengths for the non-stationary Model B. This is because the empirical heuristics report an $n_{\rm eff}$-weighted average of $\ell_r$ over all radii considered, while the GRF explicitly reports $\ell_r$ at $r_0$, with any radial variation encoded in $\alpha_r$. In the stationary case, where $\ell_r$ is independent of reference radius, these return the same value (e.g., Model A), while when $\ell_r$ varies with reference radius, $\hat{\ell}_r$ is biased towards $\ell_r$ in annuli with larger $n_{\rm eff}$.

Another notable departure is the recovery of $\ell_s$ for Model D where both approaches are biased low. As Model C, which is identical to Model D in all but the pitch, shows no such bias, we can associate this bias with the $10\degr$ tilt of the correlation kernel. As can be seen in Figure~\ref{fig:s2_atlas}m, the non-zero $\beta$ rotates the $S_2$ about the origin. As both analyses here only consider on-axis measurements (i.e., taking cuts along $\delta r = \delta \phi = 0$), the elliptical $S_2$ presents a shorter effective correlation length. To recover the input arc-length correctly, the rotation must also be considered, as described in Section~\ref{sec:numerical_simulations}.

Interestingly, we do not find a bias in the azimuthal stationarity index. All four models return $\hat{\alpha}_\phi$ within $0.23\,\sigma$ of expectation, including the three drawn at fixed arc length. The azimuthal stretching that the polar representation imposes at small radii (Section~\ref{sec:structure_function:intuition}) is thus absorbed entirely by measuring the slope on the native angular width, and does not propagate into the recovered parameters.

\section{Detection and Characterization in the Presence of Noise}
\label{sec:noisy_models}

In the previous Section we concerned ourselves only with noise-free models, Models A through D. Even without the introduction of observational noise, we identified sampling variance as an unavoidable source of `noise' and noted additionally that the polar-coordinate representation introduces an unavoidable level of non-stationarity into the field. This is absorbed by measuring $\hat\alpha_\phi$ on the native angular half-power width and so does not propagate into the recovered parameters; the residual biases that do remain --- most notably the arc length recovered for the tilted kernel of Model D --- could not be directly corrected for, but we demonstrated that by drawing many realizations of the observed field we could robustly determine the magnitude and direction of the bias, as well as quantify the variance expected from the sampling variance. In this Section, we additionally consider observational noise and extend the analysis pipeline to account for it and better quantify whether a derived heuristic is significantly different from the null expectation (i.e., the value expected from a field that was purely noise).

\subsection{Noise Models}
\label{sec:noisy_models:noise_models}

For this Section we use the same intrinsic velocity fields as described in Section~\ref{sec:morphological_models:models} but corrupted with a random noise field. For the noise, we adopt the morphology of Model A: a stationary, isotropic field with a tunable spatial scale and variance (essentially a white noise field convolved with a circular Gaussian function). We consider three spatial scales for the noise, $\ell_{0,n} = \{120,\, 60,\,30\}$~mas, equivalent to generating beams of FWHM $\{200,\, 100,\, 50\}$~mas, or, read as interferometric image noise (Section~\ref{sec:structure_function:heuristics}), to restoring beams of $\{283,\, 141,\, 71\}$~mas; the conditions below are therefore conservative relative to real data at the quoted beam sizes, whose noise would be correlated on a shorter scale and so be more readily separated from the intrinsic field, such that the noise fluctuations are, on average, either twice, equal to, or a half of the spatial scale of the intrinsic velocity fluctuations. We additionally consider three different variances, $\sigma_n = \{1,\, 0.5,\, 0.25\}$, i.e., equal to, a half of, or a quarter of the variance of the intrinsic velocity field. We do not consider cases where the noise variance is greater than the field variance. Based on the results from the exoALMA program \citep{Fukagawa_ea_2026}, these variances are likely very conservative and represent the most challenging cases where this methodology would be applied.

In this Section we only consider the case where we believe we are able to develop an accurate noise model in the projection of the data that we work with. In reality, the noise found in interferometric data is propagated through the non-linear imaging process into an image cube which is subsequently collapsed (likely involving some threshold-masking and smoothing introducing further non-linearity) into a summary statistic moment map. When applying to real data it is therefore suggested to mimic this pipeline and forward model noise through this process as accurately as possible. We have found, however, that the propagation of noise through the collapse of an image cube to a moment map is sensitive to the underlying emission morphology (i.e., whether the noise lies on top of a bright region, or there are strong velocity gradients) meaning that this becomes intractable for the goal of this work: we would need to cover a range of parameters describing the emission morphology of the disk for a single velocity draw. The following sections should therefore be considered as a demonstration of principle with the numerical results acting as a general guide as to what is possible rather than a definitive statement on the fidelity of these analyses.

\subsection{Analysis Workflow}
\label{sec:noisy_models:analysis_workflow}

An attractive property of the structure function is that it is additive such that we can decompose the structure function of our noisy data into the structure function of the noise field and a separate structure function of the velocity field. This requires us to consider additional steps in our analysis pipeline where we model the noise and then subtract it from the data.

The first step is to develop a noise model that is believed to be a good representation of the data. A reasonable starting point for ALMA-like data is to adopt the same model used in Section~\ref{sec:noisy_models:noise_models}, namely a stationary and isotropic correlated GRF. The variance can be estimated by measuring the variance of a signal-free region of the data, and the correlation length measured directly from the autocorrelation of a signal-free channel. If it is instead to be set from the synthesized beam, the appropriate value is $\theta_b / \sqrt{2}$ and not $\theta_b$ itself, for the reason given in Section~\ref{sec:structure_function:heuristics}; taking the beam width directly over-smooths the null by a factor of $\sqrt{2}$. For this work we will assume that we can extract these properties perfectly. This is a deliberate scoping choice rather than a convenience. Every significance test in this paper is therefore measured against a \emph{matched} null, drawn at the true noise amplitude, and the results that follow should be read as what is recoverable \emph{given} a correct noise model. How accurately such a model can be built from the data themselves, and how these results degrade when it is misspecified, is a separate question that we do not address here. With the statistical properties of the noise to hand, it is possible to build a null ensemble. This is many realizations of the noise which can be used to calculate a noise-only structure function, $S_2^{\rm noise}$. In order to suppress the issues associated with sampling variance we draw $N=1{,}000$ realizations and average their structure functions together.

This noise structure function can then be used to de-noise the structure function calculated from the noisy data (note that you de-noise the structure function, not the actual data). If the noise model is correct, this de-noised structure function should contain all the information regarding the structure of the underlying velocity field, $S_2^{\rm field}$. It is this de-noised structure function that we will be working with in the following sections.

Two further ensembles let us attach uncertainties and test for significance, both reusing this same machinery. To estimate the sampling variance on a measured heuristic, we model the de-noised $S_2^{\rm field}$ as a GRF (Section~\ref{sec:structure_function:parameterization}) and draw a \emph{sampling ensemble} of $N = 1{,}000$ realizations of that best-fit field, each corrupted with an independent noise draw and de-noised exactly as the observation was; the spread of the recovered heuristics across this ensemble is our estimate of their uncertainty.

To decide whether a measured heuristic is significant we compare it against a \emph{null ensemble}: $N = 1{,}000$ noise-only realizations, each de-noised with the same $S_2^{\rm noise}$. In the absence of any velocity structure this should return zero, but sampling variance and the polar-coordinate representation leave a non-zero residual whose distribution defines the null. This null is matched, drawn at the true noise level, so that it already accounts for any structure the noise itself can imprint; a heuristic is deemed significant when it lies more than three standard deviations from the null median.

\subsection{Detection}
\label{sec:noisy_models:detection}

\begin{figure*}
    \centering
    \includegraphics[width=\textwidth]{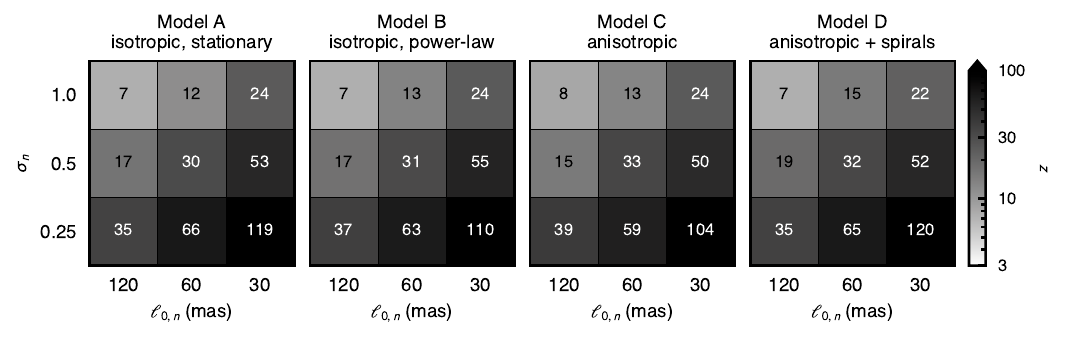}
    \caption{Significance of $\hat\sigma$ against the noise-only null, for each intrinsic velocity field (columns) over the $3\times3$ grid of noise amplitude $\sigma_n$ and spatial scale $\ell_{0,n}$. The intrinsic fields all have $\sigma = 1$. The null is matched, drawn at the true $\sigma_n$ (Section~\ref{sec:noisy_models:analysis_workflow}). Excess variance is recovered at $z > 3$ in every cell, and the significance grows as the noise is confined to smaller scales, where it shares least of its power with the broader velocity structure.}
    \label{fig:noise_significance}
\end{figure*}

The first question to ask of the data is whether there is evidence for any structure beyond what is expected from the noise model. To test for this, we first define a `null' to act as the baseline. This comprises $N=1{,}000$ noise-only realizations, generated from the noise-model appropriate for the data, and passed through the pipeline to form 1,000 $S_2$ functions which are de-noised using $S_2^{\rm noise}$, just like the data, and then analyzed using the same empirical heuristics yielding $\hat{\sigma}_{\rm null}$. The null is matched to the true noise properties, as described in Section~\ref{sec:noisy_models:analysis_workflow}. In Figure~\ref{fig:noise_significance} we plot the $z$-score, defined as
\begin{equation}
    z = \frac{\hat\sigma_{\rm obs} - {\rm median}(\hat\sigma_{\rm null})}{{\rm std}(\hat\sigma_{\rm null})},
    \label{eq:detection_z}
\end{equation}
where a value of 3 or more is taken as a significant detection of variance above the noise. Elevated variances are robustly recovered in every cell of the grid, ranging from $z \approx 7$, where the noise equals the field in amplitude and has twice its correlation length, to $z \approx 120$, where the noise is a quarter of the amplitude on half the scale. The significance grows both as the noise amplitude falls and as the noise is confined to smaller spatial scales, where it shares least of its power with the broader velocity structure. Detecting excess variance is therefore not the limiting step; the harder question, addressed next, is whether the structure of that variance can be characterized.

\subsection{Characterization}
\label{sec:noisy_models:characterization}

\begin{figure*}
    \centering
    \includegraphics[width=\textwidth]{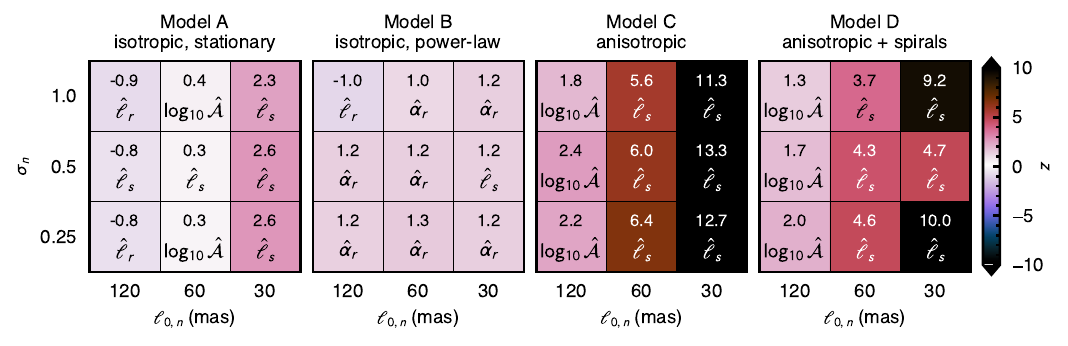}
    \caption{For each velocity model (columns) and noise condition (a $3\times3$ grid over noise amplitude $\sigma_n$ and spatial scale $\ell_{0,n}$), the signed significance $z$ of the structural heuristic that departs most strongly from the matched noise-only null. The label beneath each value names the heuristic responsible. The color scale is linear in $z$ and saturates at $|z| = 10$; the dashed marks on the color bar are the $|z| = 3$ significance level, and near-white cells within them are consistent with noise. The amplitude $\hat\sigma$ is excluded here as it is the subject of Figure~\ref{fig:noise_significance}.}
    \label{fig:significance_maxz}
\end{figure*}

A natural follow-up question is whether we can characterize the structure of the residuals now that the enhanced variance is robustly detected. We apply the same significance test to the five structural heuristics $\{\hat{\ell}_r,\, \hat{\ell}_s,\, \log_{10}\hat{\mathcal{A}},\, \hat{\alpha}_r,\, \hat{\alpha}_\phi\}$. The significance is defined slightly differently here than in Equation~\ref{eq:detection_z}. For detection the relevant comparison is whether the observed amplitude lies outside the null distribution, so the null spread alone sets the scale. For characterization we ask instead whether the observed and null values are distinguishable given both sources of uncertainty, and so we combine the sampling spread of the model-plus-noise ensemble with the spread of the null in quadrature. Both spreads are taken as $16$--$84$th percentile half-widths, and where the distributions are asymmetric we use the side facing the other distribution, making the test the more conservative of the two available choices. For the anisotropy the percentiles are taken on $\log_{10}\hat{\mathcal{A}}$ itself rather than propagated from the linear ratio, so that the test cannot depend on which way round the ratio is written. Figure~\ref{fig:significance_maxz} shows, for each model and noise condition, the signed $z$-score of whichever heuristic departs most strongly from that null, labeled by the heuristic responsible. The $z$-scores for the individual heuristics are given in Appendix~\ref{sec:appendix_significance}.

The azimuthal correlation length $\hat{\ell}_s$ carries the detection wherever there is one. Of the $36$ cells, $12$ reach $|z| \geq 3$; all $12$ belong to the anisotropic Models C and D, and in every one of them $\hat{\ell}_s$ is the heuristic responsible. Its significance climbs steeply as the noise is confined to smaller scales, but falls to $|z| \approx 2$ or below once the noise is twice the scale of the field. The anisotropy is the only other heuristic to reach significance, and it does so weakly and only for the zero-pitch Model C. $\log_{10}\hat{\mathcal{A}}$ is the strongest heuristic in eight cells, six of them at the largest noise scale where $\hat{\ell}_s$ has lost its own significance. Its sign is the expected one throughout, $\hat{\mathcal{A}} > 1$ for these azimuthally elongated fields. The stationarity slopes $\hat{\alpha}_r$ and $\hat{\alpha}_\phi$ never exceed $|z| \approx 1.3$. What is recoverable in this regime is therefore the azimuthal correlation length of an anisotropic field and, marginally and only where that anisotropy is strong and non-zero, the anisotropy itself. The radial variation of either length is not.

\subsection{Parameter Recovery}
\label{sec:noisy_models:parameter_recovery}

\begin{figure*}
    \centering
    \includegraphics[width=\textwidth]{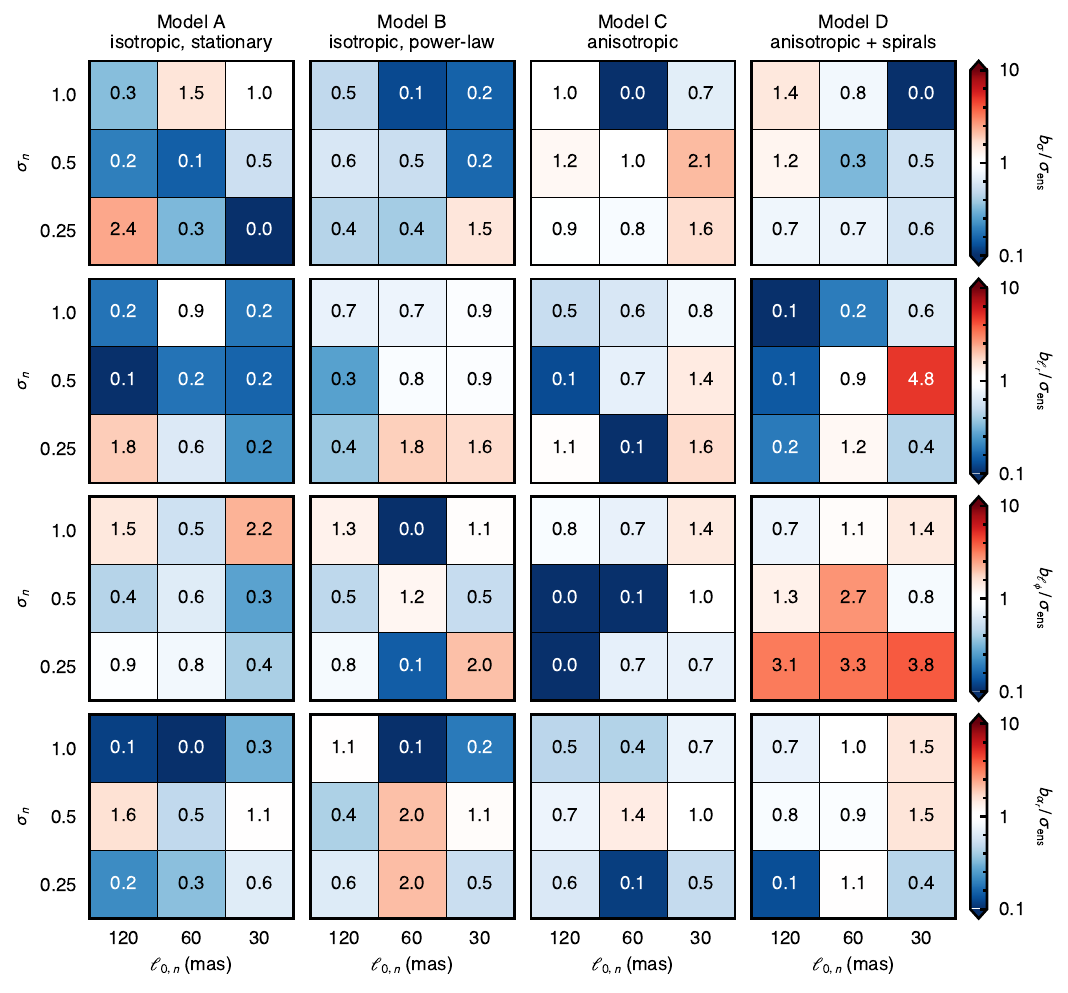}
    \caption{Demonstrating the bias in the model properties in units of the spread from the variance. Values which are less than one indicate that the bias is smaller than the uncertainty. Each column corresponds to a different GRF model (A through D), while each row shows the result for a different model parameter: $\sigma$, $\ell_r$, $\ell_\phi$ and $\alpha_r$, top to bottom. Every panel displays a 3x3 grid of noise properties, the x-axis varying the spatial scale of the noise and the y-axis varying the variance. The dashed line on each color bar marks the unity ratio.}
    \label{fig:param_recovery_ratio}
\end{figure*}

Given that simple heuristics are able to recover some structure in the de-noised models, we explore in this section how well we can extract properties of the noise. Figure~\ref{fig:param_recovery_ratio} quantifies how accurately the parametric fit recovers the input parameters in the presence of noise. For each parameter we plot the ratio of the bias, the median recovered value minus the truth, to the ensemble spread $\sigma_{\rm ens}$ (the $16$--$84$th percentile half-width of the sampling ensemble). A ratio below unity (blue) means the systematic offset is smaller than the statistical uncertainty and the parameter is recovered consistently, while a ratio above unity (red) flags a systematic bias.

The radial correlation length $\ell_{0,r}$ is the best-recovered parameter, with a bias that stays inside the sampling scatter in $28$ of the $36$ cells. The exceptions sit mostly at the lowest noise amplitude, where the ensemble scatter shrinks faster than the residual offset does. The azimuthal length $\ell_{0,\phi}$ is the least reliably recovered of the four. Its fractional scatter matches that of $\ell_{0,r}$, but its median fractional bias is roughly twice as large. It is unbiased in the median for Models A and C over most of the grid, and it is low for Model~D, attributable mainly to the $10\degr$ pitch rather than to the noise. The amplitude $\sigma$ carries no systematic bias, but its cell-to-cell scatter is larger than $N = 1{,}000$ realizations might suggest. The limitation on $\sigma$ is therefore the sampling variance rather than a consistent bias. The radial scaling $\alpha_r$ is recovered only when it is genuinely present. For the sloped Model~B ($\alpha_r = 1.0$) it is essentially unbiased, whereas for the nominally stationary models it wanders as the noise mimics a rising or a declining structure function. A fitted slope is thus a reliable indicator of genuine radial scaling but not a sensitive test of stationarity.

\begin{figure*}
    \centering
    \includegraphics[width=\textwidth]{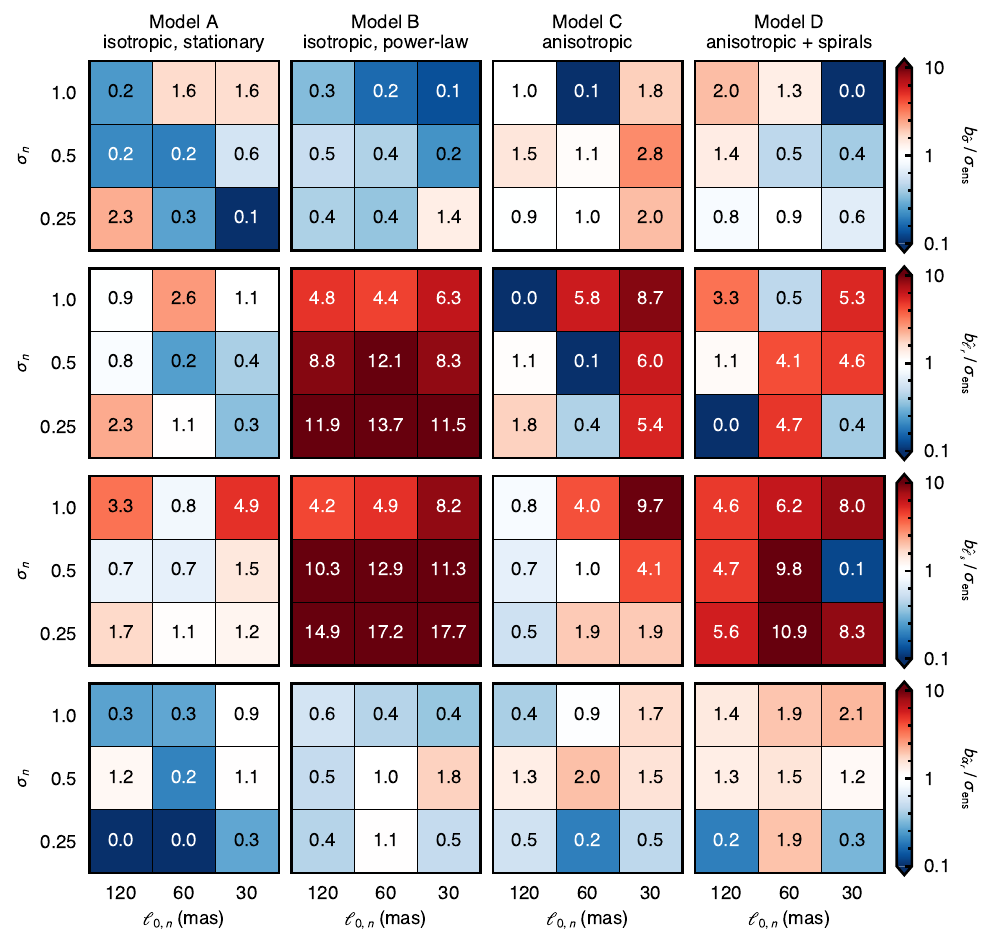}
    \caption{As Figure~\ref{fig:param_recovery_ratio}, but for the empirical heuristics $\{\hat\sigma,\, \hat\ell_r,\, \hat\ell_s,\, \hat\alpha_r\}$ rather than the parametric fit -- the noisy-regime analogue of the empirical (gray) points in Figure~\ref{fig:model_heuristics_vs_grf}. Reading the two figures together shows which estimator, empirical or parametric, degrades more slowly with noise.}
    \label{fig:param_recovery_ratio_heuristics}
\end{figure*}

The parametric fit results are the estimator one would ultimately quote as a measurement, but these are not the only option. Figure~\ref{fig:param_recovery_ratio_heuristics} repeats the recovery test for the empirical heuristics, providing the noisy-regime counterpart to the noise-free comparison of Figure~\ref{fig:model_heuristics_vs_grf}. Which estimator suffers more from the noise is parameter-dependent, but the fit is the better of the two more often than not: across the four parameters and nine noise conditions of the four models, the heuristic bias exceeds its own ensemble scatter in $83$ of $144$ cells, against $41$ for the fit. For the amplitude the two are close, as they were in the absence of noise. For the correlation lengths the fit is decisively better: its bias exceeds its scatter in $8$ of $36$ cells on $\ell_{0,r}$ and $14$ on $\ell_{0,\phi}$, against $25$ and $29$ for the heuristics. When the heuristics fail, they appear to fail badly (up to $18\sigma$ against a worst case scenario of $4.8\sigma$ for the fit). Much of that gap, however, is the definitional difference identified in Section~\ref{sec:structure_function:models} rather than a sensitivity to noise and is accordingly widest for the non-stationary Model~B. For the radial scaling the fit approach performs better, though both are only marginally constrained. The takeaway is that both methods robustly recover the amplitude of the variations, but measures of the correlation lengths are more accurately reproduced with the GRF fit.

\section{(Magneto-)Hydrodynamical Turbulence}
\label{sec:numerical_simulations}

Morphological models are an efficient way to test the analysis pipeline and demonstrate that we are able to characterize structure in the noise-dominated regime. In this section, we extend the question posed in \citet{Barraza-Alfaro_ea_2025} regarding the identification and differentiability of three (magneto-)hydrodynamical (MHD) instabilities, as represented by one simulation of each, namely the vertical shear instability (VSI), the gravitational instability (GI) and the magneto-rotational instability (MRI). These three instabilities are known to drive qualitatively different structures: the VSI generates concentric, annular features \citep{Nelson_ea_2013}; the GI is synonymous with nested, grand-design spirals \citep{Hall_ea_2020}; and the MRI drives flocculent spiral arms \citep{Flock_ea_2015}.

\begin{figure*}
    \centering
    \includegraphics[width=\textwidth]{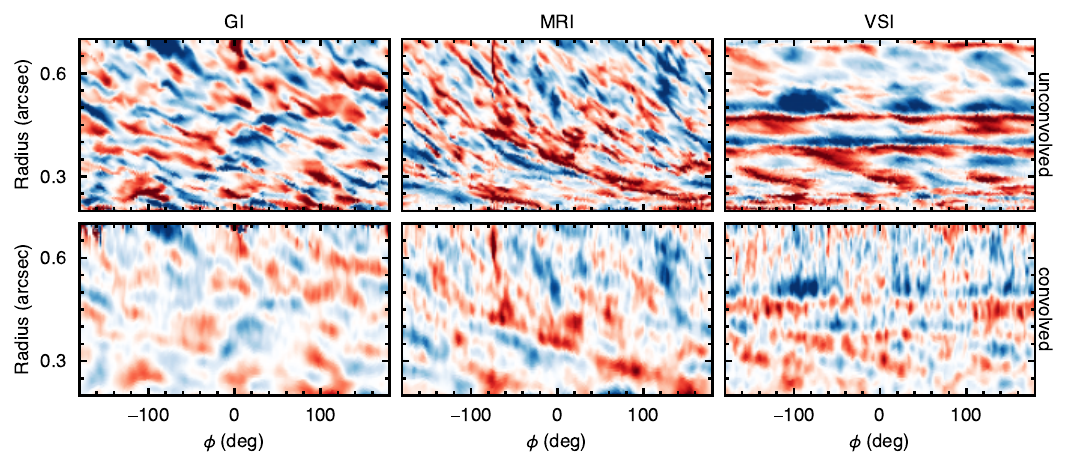}
    \caption{Polar projections of the variance-normalized line-of-sight velocity residuals from numerical simulations of the GI, left, MRI, center, and VSI, right, extracted from synthetic image cubes of the face-on disks with the \texttt{percentiles} method in \texttt{bettermoments} after subtracting a Keplerian component. The top row shows the `raw' maps, extracted from cubes without noise or beam convolution, while the bottom row shows the `convolved' maps, extracted from cubes with added noise and a beam convolution designed to mimic the quality of the exoALMA observations as in \citet{Barraza-Alfaro_ea_2025}.}
    \label{fig:polar_plots_simulations}
\end{figure*}

We use the same three simulations presented in \citet{Barraza-Alfaro_ea_2025}, and shown in Figure~\ref{fig:polar_plots_simulations}, referring the reader to that work for more details of these, and use the face-on ($i = 0\degr$) line-of-sight velocity maps, extracted from the synthetic image cubes using the \texttt{percentiles} method in \texttt{bettermoments} \citep{Teague_Foreman-Mackey_2018}, from which a Keplerian component is subtracted to leave the velocity residuals. For a face-on disk the line of sight is aligned with the rotation axis, so these residuals trace the vertical velocity perturbations rather than the rotation. We focus on the face-on disks to increase the interpretability of the results, using these as a demonstration of the analysis technique rather than an exhaustive scan of all observing conditions possible. In this paper we do not consider planet-disk interactions as those kinematic structures are generally localized and thus do not benefit from the full-field context of the $S_2$ \citep[instead they are far more detectable through feature-detection techniques such as those described in][]{Izquierdo_ea_2026}.

Both a `raw' and a `convolved' map are considered and were derived from image cubes imaged with $104~{\rm m\,s^{-1}}$ channel spacing. The `raw' velocity map is produced in the absence of noise and any beam convolution, while the `convolved' map is processed, at the level of the image cube and prior to the velocity extraction, to mimic exoALMA-quality data \citep{Teague_ea_2025}, as described in \citet{Barraza-Alfaro_ea_2025}. This includes convolution with a circular 35~mas beam (a smaller beam than used in exoALMA because the goal is to conserve the number of beams-per-radius rather than match linear scales) and the injection of noise with an rms of $\sigma \approx 0.16~{\rm mJy\,beam^{-1}}$. This results in noise which is sufficiently well described by an isotropic GRF with a standard deviation of $\sigma \approx 10~{\rm m\,s^{-1}}$. As noted in Section~\ref{sec:noisy_models}, this is only an approximation of the noise propagation through the moment-map creation process and so should be considered as a demonstration of the technique rather than a definitive application.

\subsection{Detection of Structure}
\label{sec:numerical_simulations:detection}

We first test whether our analysis pipeline is able to detect the presence of structure in the `convolved' data. Following Section~\ref{sec:noisy_models:detection}, this involves developing a noise model that allows us to both de-noise the structure function of the observation and generate a null ensemble with which to compare our extracted values. As we only have a single observation, we are unable to model the scatter from the variance, but we can use the results from Section~\ref{sec:noisy_models} as a guide.

For the noise model we adopt the injected noise directly: a flat $\sigma_n = 10~{\rm m\,s^{-1}}$ at every radius, with each realization drawn as white noise convolved with a Gaussian kernel of FWHM $35/\sqrt{2} \approx 25$~mas, so that the noise is correlated on the 35~mas beam itself. $N=1{,}000$ realizations were drawn and the average of their $S_2$ taken as the de-noising $S_2^{\rm noise}$. An additional $N=1{,}000$ realizations were drawn to serve as the null ensemble. We find a $z$-score for $\hat{\sigma}$ of $266$, $40$ and $27$ for the GI, MRI and VSI, respectively. Structure is thus detected decisively in all three cases, which is unsurprising given that it is readily identifiable by eye in the bottom row of Figure~\ref{fig:polar_plots_simulations}. As in Section~\ref{sec:noisy_models:detection}, detection is not the limiting step; the informative question is whether the three simulations can be told apart, which we address next.

\subsection{Characterization and Differentiation of the Simulations}
\label{sec:numerical_simulations:differentiation}

\begin{figure}
    \centering
    \includegraphics[width=\columnwidth]{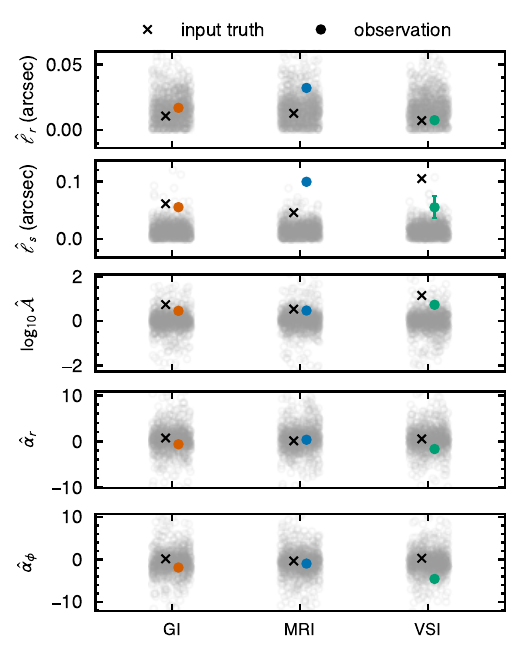}
    \caption{The five shape heuristics recovered for the three simulations, measured over the common $0\farcs2 - 0\farcs7$ band. Gray points are the noise-only null models, the colored point is the de-noised convolved pseudo-observation with its sampling-ensemble spread, and the black cross marks the value extracted from the unconvolved velocity map.}
    \label{fig:numerical_simulations_heuristic_characterization}
\end{figure}

\begin{figure}
    \centering
    \includegraphics[width=\columnwidth]{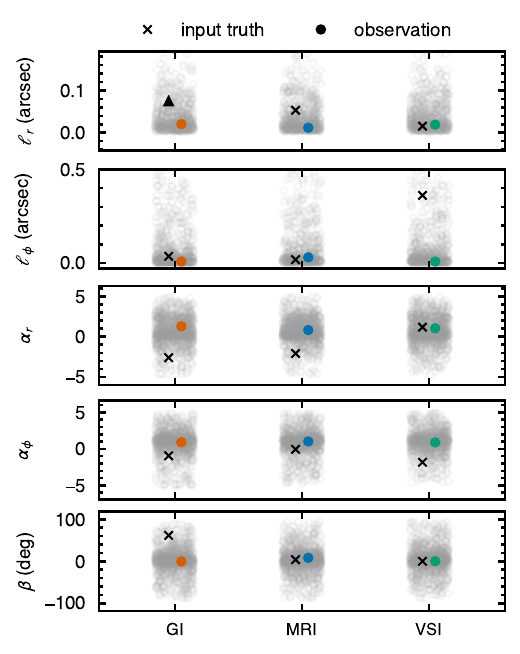}
    \caption{As Figure~\ref{fig:numerical_simulations_heuristic_characterization}, but for a parametric fit to the global-mode structure function, with the pitch angle $\beta$ freed. Gray points are the noise-only null models, the colored points the fit to the (de-noised) convolved observation, and the black cross the fit to the unconvolved model. The fitted parameters largely fall within the null cloud, and the pitch $\beta$ in particular is unconstrained for all three models (Section~\ref{sec:numerical_simulations:GRF_limitations}). To ease comparison, the correlation lengths are evaluated at $r_0 = 0\farcs37$, the logarithmic centre of the band over which Figure~\ref{fig:numerical_simulations_heuristic_characterization} measures its heuristics. Noise-only optimizations that stop on, or within $5\%$ of, a bound are not shown as they are considered a poor fit. The triangle denotes that $\ell_r$ as measured for the GI is unconstrained and hits the prior boundary.}
    \label{fig:numerical_simulations_GRF_characterization}
\end{figure}

With a clear detection possible, we then considered whether structure beyond the noise could be characterized. The noise model is unchanged from Section~\ref{sec:numerical_simulations:detection}: it is drawn at the injected amplitude, so the characterization is measured against the same matched null as the detection. This represents the case where the noise model is well constrained.

Figure~\ref{fig:numerical_simulations_heuristic_characterization} shows the empirical heuristics we derive from the `convolved' velocity map (shown in color), compared to the null models, shown as a cloud of gray points, and the value derived from the `raw' velocity map as a black cross. Figure~\ref{fig:numerical_simulations_GRF_characterization} shows the equivalent plot but with parameters from a least-squares fit of the GRF model to the global-mode structure function, which additionally constrains the pitch angle $\beta$ of any flocculent spiral. Each gray point is one such fit to a noise-only de-noised realization. In this least-squares view the GRF shape parameters largely fall within the broad noise-only null, leaving the amplitude $\sigma$ (not shown) as the only strongly detected parameter. This comparison shows that none of the GRF shape parameters separates the three simulations. We therefore take the empirical heuristics of Figure~\ref{fig:numerical_simulations_heuristic_characterization}, not the GRF fit, as our primary characterization.

These figures show that, as for the morphological models explored previously, the beam convolution and noise injection bias the recovered correlation lengths. The bias is not uniform, however, but depends on the underlying structure. This is clearest in the arc-length scale $\hat{\ell}_s$ (second row of Figure~\ref{fig:numerical_simulations_heuristic_characterization}): for the GI it is recovered essentially unbiased because the grand-design spiral is broad compared with the beam; for the VSI it is strongly suppressed as the 35~mas beam smears together the closely-spaced corrugation rings; and for the MRI it is instead inflated as the beam blends its fine filaments into broader apparent correlations. The anisotropy behaves similarly: $\log_{10}\hat{\mathcal{A}}$ sits about $1\sigma$ above the null for all three models, signaling azimuthally elongated structure, but below the noise-free truth, so the elongation is detected and yet compressed by the observation. This is most severe for the VSI. The fitted GRF pitch angle $\beta$, by contrast, is not a useful discriminator. For all three models it lies within the noise-only null ($\sigma_{\rm null} \approx 30\degr$), and in the convolved data $|\beta| < 10\degr$ throughout. The only nominally large value, the unconvolved GI fit is not a measurement of its grand-design spiral but one branch of a reflection degeneracy in the near-isotropic fluctuation field, a limitation of the GRF description examined in Section~\ref{sec:numerical_simulations:GRF_limitations}.

Read alongside Figure~\ref{fig:numerical_simulations_heuristic_characterization}, the fitted lengths in Figure~\ref{fig:numerical_simulations_GRF_characterization} contradict the heuristics. The global-mode fit returns $\mathcal{A} = \ell_\phi/\ell_r < 1$ for the GI and VSI and $> 1$ for the MRI, whereas $\log_{10}\hat{\mathcal{A}}$ places all three in the azimuthally elongated regime. The fitted ratio is not, however, a measurement, as the measured values fall squarely in the null cloud. The breadth arises from the degeneracy introduced by $\beta$ described in Section~\ref{sec:structure_function:models}. Figure~\ref{fig:numerical_simulations_GRF_characterization} therefore constrains only $\sigma$, and the anisotropy of these fields is measured by the heuristics alone.

Beyond characterizing each field individually, we can ask whether the three simulations are distinguishable from one another on the basis of these heuristics. Treating the shape heuristics $\{\hat{\ell}_r,\, \hat{\ell}_s,\, \hat{\mathcal{A}},\, \hat{\alpha}_r,\, \hat{\alpha}_\phi\}$, with the anisotropy entering linearly, as a vector and combining the differences between two models in quadrature, with each heuristic's uncertainty taken as the scatter of the null ensemble and a radial bootstrap over the reference annuli added in quadrature, we find that the de-noised observations separate in all three pairs, at $8.4\sigma$ (GI--MRI), $8.7\sigma$ (MRI--VSI) and $4.1\sigma$ (GI--VSI). The MRI simulation is therefore cleanly distinguished from both hydrodynamic simulations, whereas the grand-design GI spiral and the corrugated VSI rings are the hardest pair to tell apart at this data quality, once the beam has degraded both toward broad, low-contrast structure. We also find that this pair is also the most sensitive to the noise model. Repeating this procedure but de-noising with a null with a $\sqrt{2}$ larger beam lowers the GI--VSI separation to $2.8\sigma$, while the other two pairs remain above $6\sigma$.

These separations apply to these three velocity fields, under the adopted physical parameters and observing conditions. They do not establish that the GI, MRI and VSI occupy distinct regions of heuristic space in general. With a single snapshot per mechanism, neither the sampling ensemble nor the radial bootstrap measures the variation within a class, and different physical parameters (e.g., the disk-to-star mass ratio and cooling time for the GI, the ionization state and field strength for the MRI, or the thermal relaxation time for the VSI), numerical setups or snapshots could introduce substantial variation within each. Distinguishing the mechanisms themselves would require repeating this analysis on a suite of simulations spanning that variation.

\subsection{Limitations of the GRF Model}
\label{sec:numerical_simulations:GRF_limitations}

The experiment above also exposes a structural limitation of the GRF description. Because the Gaussian-envelope structure function is monotonically increasing and saturating, it cannot reproduce an \emph{anti}-correlation, which manifests as a peak in $S_2$ that rises above the plateau (excess variance concentrated at a particular scale). Fitting the model to the noise-free fields makes this concrete: the fractional residual the fit cannot remove is smallest for the MRI ($0.33$) and largest for the VSI ($0.51$), with the GI intermediate ($0.44$), so the MRI, whose turbulent velocity structure appears close to monotonic, is the only one of the three that the GRF genuinely describes. The radial structure functions of the GI and VSI instead overshoot their plateaus, the signature of the coherent grand-design spiral and the quasi-periodic corrugation rings, respectively. Freeing the pitch angle does not improve the fit. A pitch term only rotates a non-negative correlation envelope and can never induce the overshoot. A dedicated model, such as a coherent winding-ridge fit to the off-diagonal, global-mode 2D structure function, is the appropriate tool for these features. Where such a model is not warranted, the empirical heuristics, which collapse the same information onto the diagonal, still provide a robust first-order description of the field, as the differentiation of Section~\ref{sec:numerical_simulations:differentiation} demonstrates.

The pitch angle itself deserves particular caution. Its only signature in the structure function is the tilt of the correlation ellipse, whose amplitude scales as $\ell_\phi^2 - \ell_r^2$ and therefore vanishes as the field approaches isotropy. Injection--recovery tests on synthetic pitched fields confirm this. A known $\beta = 30\degr$ is recovered only when the field is strongly anisotropic ($\ell_\phi/\ell_r \gtrsim 4$), whereas at $\ell_\phi/\ell_r \lesssim 2$ the recovered value is uninformative (scatter $\gtrsim 30\degr$). The fluctuation fields here are near-isotropic or at best marginally anisotropic. No field in either condition, either raw or convolved, approaches the threshold, so $\beta$ is intrinsically unconstrained throughout. Compounding this, the kernel admits an exact reflection degeneracy where $(\ell_r,\ell_\phi,\beta)$ and $(\ell_\phi,\ell_r,\beta\pm90\degr)$ produce identical structure functions meaning that a least-squares fit returns one arbitrary branch with a formal error that significantly understates the true uncertainty.

Finally, a grand-design spiral such as the GI is coherent. Its arm is a deterministic, phase-locked feature carried in the mean field, whereas the pitched GRF is a flocculent, random-phase model whose pitch describes only the anisotropy of the fluctuations about that mean. The flocculent pitch cannot capture a coherent arm even in principle, reinforcing that the coherent winding-ridge model is the correct tool. For real data the practical lessons are to gate any pitch measurement on the measured anisotropy, to treat the formal GRF parameter uncertainties skeptically wherever the fit is misspecified (a large fractional residual), and to lead with the empirical heuristics, which remain robust where the parametric fit does not.

\section{Application to TW Hya}
\label{sec:twhya}

\begin{figure*}
    \centering
    \includegraphics[]{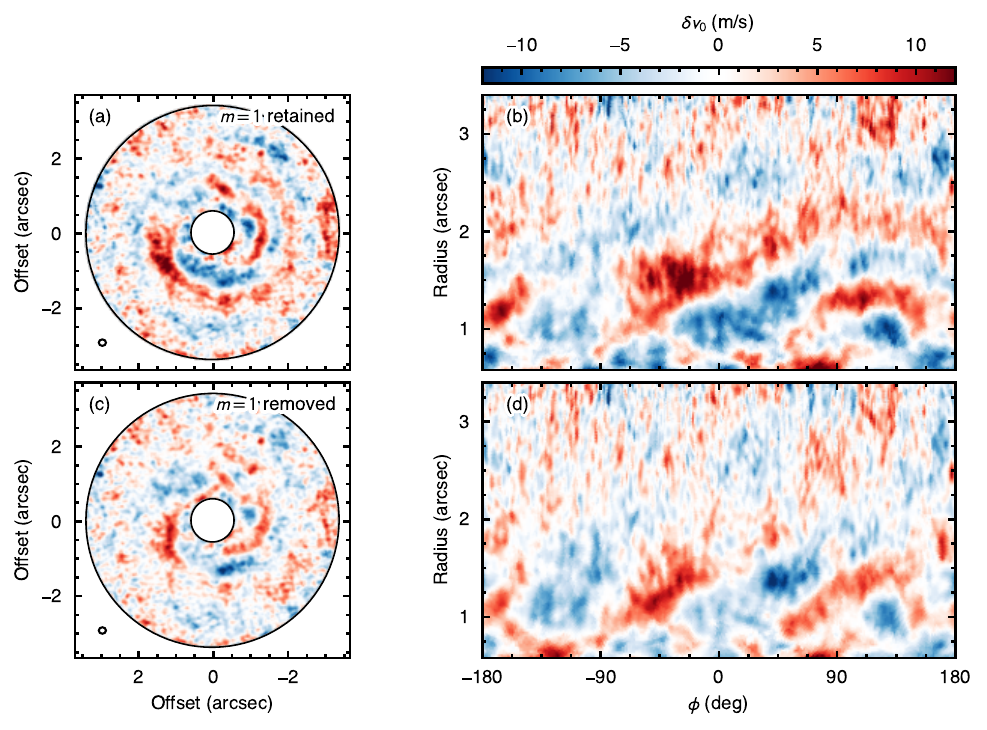}
    \caption{On-sky and polar projections of the velocity perturbations in TW~Hya traced by $^{12}$CO emission. The top row shows the case when a Keplerian rotation model including pressure support is subtracted from the observed velocity field \citep[as done in][]{Teague_ea_2022}, while the bottom row shows where the $m=1$ component is removed through an annulus-by-annulus modeling approach. The beam is shown in the lower left of each on-sky projection.}
    \label{fig:polar_plots_twhya}
\end{figure*}

The preceding sections establish what the structure function can recover under controlled conditions. Here we turn to real data to illustrate what an observer stands to learn in practice. TW~Hya is a natural first target given that it is viewed almost face-on, removing any ambiguity from backside emission. It is also known to exhibit one prominent large-scale feature, an $m=1$ spiral, on top of residuals that otherwise appear noise-like by eye, as shown in Figure~\ref{fig:polar_plots_twhya}. With the structure function analysis we seek to ask four questions of the data: is there correlated structure beyond the noise; on what radial and azimuthal scales is it correlated; how do those scales change with radius; and how much of it is carried by the feature visible by eye?

We use observations of CO from TW~Hya, originally presented in \citet{Teague_ea_2022}. The images were made with a channel spacing of $40~{\rm m\,s^{-1}}$, the spectral resolution of the data, and restored with a beam of $0\farcs 19 \times 0\farcs 17$ $(89\degr)$. The data were corrected with the `JvM-correction', as described in \citet{Jorsater_vanMoorsel_1995} and \citet{Czekala_ea_2021}, resulting in noise levels of $0.6~{\rm mJy~beam^{-1}}$. While more recent kinematic studies have opted not to apply the `JvM-correction', several works have shown that the inclusion of this correction has little impact on the kinematic signals in the data \citep[e.g.,][]{Woelfer_ea_2026}. We refer the reader to the original paper for more information regarding observations, calibration and imaging of the data.

Unlike the original paper, we calculate the velocity maps using the percentile method in \texttt{bettermoments}, as used to make the maps used in Section~\ref{sec:numerical_simulations}. We consider two different models of the background velocity structure. The first is the same as performed in \citet{Teague_ea_2022} where a Keplerian model including a retardation due to a negative pressure gradient is fit to the data. The second removes $m=1$ structure by modeling the background velocity on an annulus-by-annulus basis where azimuthally symmetric $v_\phi$, $v_r$ and $v_z$ components are considered. This annulus-by-annulus approach is similar to the approach described in \citet{Casassus_Perez_2019}, however the disk inclination and position angle are kept constant across all annuli with $i = 5\fdg8$ and ${\rm PA} = 151\fdg6$. We consider this subtraction as it removes the large-scale $m=1$ spiral feature which is poorly captured by GRFs (and which \citealt{Winter_ea_2025} demonstrate may be due to a warped disk, rather than the planet-driven buoyancy spirals originally proposed by \citealt{Bae_ea_2021}), while also accounting for the sub-Keplerian rotation in the outer disk attributed to large pressure gradients. The resulting on-sky and polar projections of the velocity fields are shown in Figure~\ref{fig:polar_plots_twhya}.

\begin{figure}
    \centering
    \includegraphics[width=\columnwidth]{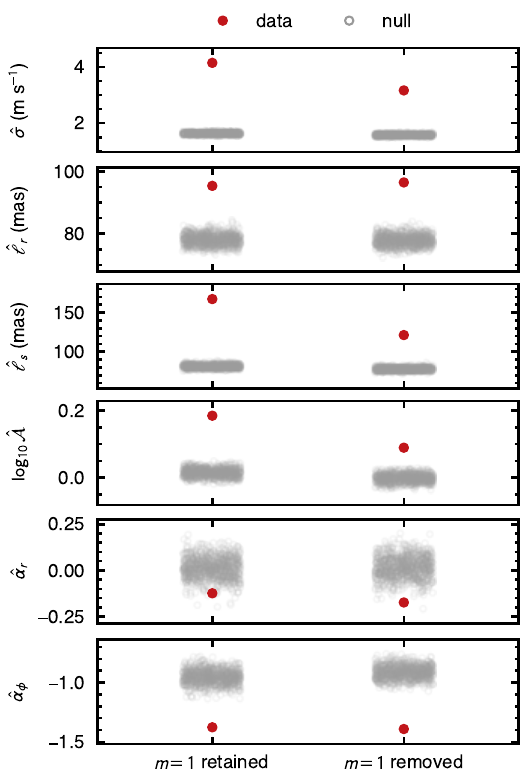}
    \caption{Empirical heuristics of the velocity field traced by CO in TW Hya considering when the $m=1$ spiral is preserved or removed. Both cases recover the same trend for all parameters. The `$m=1$ retained' case departs further from the null ensemble in $\hat\sigma$, $\hat\ell_s$ and $\hat{\mathcal{A}}$, while the two cases are comparably significant in $\hat\ell_r$, $\hat\alpha_r$ and $\hat\alpha_\phi$. Typical values are $\hat\ell_r \sim 95~{\rm mas}$ (6~au), $\hat\ell_s \sim 120 - 170~{\rm mas}$ ($7 - 10$~au), $\hat\alpha_r \sim -0.15$ and $\hat\alpha_\phi \sim -1.4$, such that $\hat{\mathcal{A}} \propto r^{\hat\alpha_\phi - \hat\alpha_r + 1} \approx r^{-1/4}$.}
    \label{fig:twhya_heuristics_nulls}
\end{figure}

We apply our structure function analysis to both cases between $0\farcs58$ (three times the beam major axis) and $3\farcs4$. Beyond $3\farcs4$ the CO emission fades steeply towards the edge of the disk, such that the noise in the velocity map grows with radius, a trend that the uniform-amplitude null described below does not capture. We report only the empirical heuristics as the parametric GRF fit is not robust for these data as discussed in Section~\ref{sec:numerical_simulations:GRF_limitations}. Figure~\ref{fig:twhya_heuristics_nulls} displays the extracted heuristics for both cases compared to a null ensemble.

This application to real data demonstrates that the null must be forward-modeled as we find that the formal uncertainties returned by \texttt{bettermoments} are not a reliable measure of the pixel-to-pixel noise. Beyond $\sim 1\farcs5$ they exceed the rms of the full residual map, signal included, by a factor of two or more, so a null scaled to them would place the observed variance below the noise. We instead drew a separate null ensemble for each case from white noise convolved with a Gaussian kernel of FWHM $\theta_b / \sqrt{2}$, so that the noise is correlated on the restoring beam itself and scaled it to a per-pixel amplitude measured from noise-only realizations of the image cube passed through the same \texttt{bettermoments} pipeline.

Figure~\ref{fig:twhya_heuristics_nulls} presents the returned empirical heuristics which we interpret in the context of the four questions previously posed. First, correlated structure is significantly detected for both background models ($\hat\sigma$ is robustly detected relative to the null sample). Second, its scales differ sharply between the two directions. Radially, $\hat\ell_r \sim 95~{\rm mas}$ (6~au at a distance of 60.1~pc) lies only slightly above its null value, and $\hat\alpha_r \approx 0$ shows that it is constant with radius, as expected if the beam rather than the disk sets it. The radial width of the perturbations are no wider than $\sim 6$~au, smaller than the local gas scale height over most of the radii considered. Azimuthally, by contrast, the structure is well resolved and coherent along arcs spanning several beams. Third, the arc length shrinks outwards, so the elongation weakens with radius, $\hat{\mathcal{A}} \propto r^{-1/4}$, broadly as expected for spiral arms that open up at larger radii, a feature that can be tentatively seen in Figure~\ref{fig:polar_plots_twhya}d. Fourth, comparing the two background models isolates the $m=1$ feature. Differencing $\hat\sigma$ for the two in quadrature, the spiral carries an rms of $\approx 2.7~{\rm m\,s^{-1}}$, roughly half of the variance above the noise in the retained map. It contributes amplitude and azimuthal coherence, but not the outward shrinking of the arc length: its removal shortens $\hat\ell_s$ while leaving $\hat\ell_r$ and $\hat\alpha_\phi$ essentially unchanged. The structure that remains after its removal is still significantly elongated azimuthally, and TW~Hya thus hosts azimuthally coherent velocity perturbations beyond the feature visible by eye.

This example demonstrates the utility of such a statistical analysis. Even where the residuals look noise-like, it decides whether correlated structure is present, quantifies its amplitude and measures how far it is coherent in azimuth. It is also powerful in that it clearly demonstrates what the data cannot constrain. As found for the synthetic fields of Section~\ref{sec:noisy_models}, the radial scale is set by the beam and yields only an upper limit. Finally, the amplitude and azimuthal heuristics depend on the adopted background model and must be reported alongside it, whereas the radial heuristics, being resolution-limited, are largely insensitive to it.

\section{Beyond Face-On Disks}
\label{sec:inclined}

Every demonstration in this paper uses a face-on, or very nearly face-on, disk, meaning that the transformation from an on-sky projection to one in polar disk coordinates is greatly simplified relative to mid-inclination disks. Furthermore, with no contamination from the back side of the disk \citep[which results in asymmetric or double-peaked line profiles;][]{Izquierdo_ea_2025}, we can be confident that the fluctuations we identify are real velocity fluctuations rather than systematic features associated with poorly constrained geometrical properties. As the majority of sources are viewed at moderately inclined projections, we discuss how the application and interpretation of this methodology would change when considering more inclined sources.

\subsection{Source Geometry}
\label{sec:inclined:geometry}

A disk which is viewed inclined relative to the line of sight will result in the beam unevenly sampling the disk: the major axis becomes less well sampled azimuthally, while the minor axis is poorly sampled along the radial direction resulting in an effective resolution that changes with $(r,\, \phi)$. Furthermore, the reduction in projected disk area results in a factor $\sim \cos i$ fewer independent beams at a given azimuth, increasing the sampling scatter in $S_2$.

To test how these effects change the heuristics, we repeated the analysis of Section~\ref{sec:noisy_models} for Models A and C at $i = 0\degr$, $30\degr$ and $60\degr$. Each realization was projected onto the sky, convolved with a circular 50~mas beam, corrupted with noise of $\sigma_n = 0.5$ drawn in the sky plane and correlated on the 50~mas beam, and deprojected onto the same polar grid as before. While this assumes a geometrically thin disk with no flaring, the setup still provides insight as to the dominant impact of elevated inclinations. The noise used for de-noising and for the null ensemble was drawn on the sky and passed through the same deprojection, so that the null carries the inclined geometry.

\begin{figure}
    \includegraphics[]{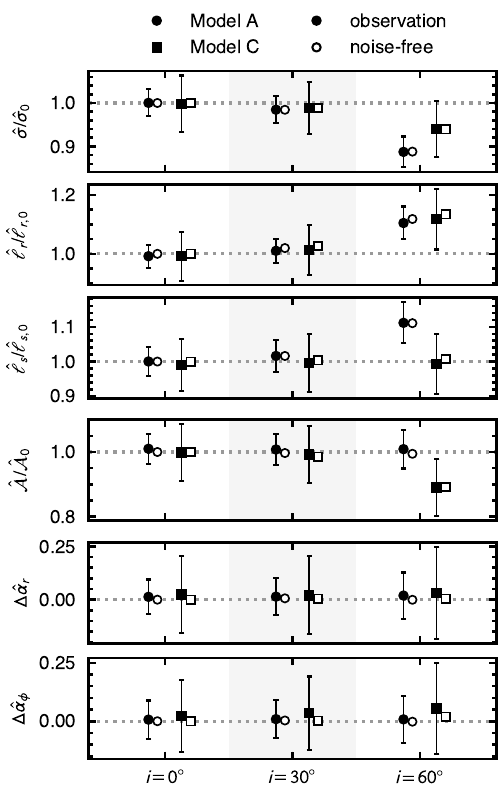}
    \caption{Heuristics recovered for Models A (circles) and C (squares) at inclinations of $0\degr$, $30\degr$ and $60\degr$, relative to the noise-free, beam-convolved field of the same model viewed face-on (subscript 0). Filled markers show the de-noised observation, with the error bars signifying the spread across the sampling ensemble, and open markers the noise-free convolved field at that inclination. Only at large inclinations of $i = 60\degr$ are the heuristics strongly biased.}
    \label{fig:inclination_recovery}
\end{figure}

Figure~\ref{fig:inclination_recovery} compares each heuristic with its value for the same model viewed face-on. At $30\degr$ every heuristic is recovered within a few per cent of its face-on value. At $60\degr$ the stretched beam smooths the field, lowering $\hat{\sigma}$ and lengthening $\hat{\ell}_r$ by roughly ten per cent; for Model A, $\hat{\ell}_s$ grows by a similar amount and the field remains isotropic, while for Model C the long azimuthal correlation is barely affected but its anisotropy $\hat{\mathcal{A}}$ is compressed. The slopes $\hat{\alpha}_r$ and $\hat{\alpha}_\phi$ remain consistent with the null at every inclination. Detection remains robust across all inclinations, with the significance falling roughly as expected from the reduced number of independent beams.

We note that at every inclination the de-noised heuristics match those of the noise-free convolved fields (compare the circle to square markers). This demonstrates that the noise introduces no inclination-dependent bias once the null is drawn through the same deprojection. The geometry of an inclined source is therefore absorbed by the matched-null framework for the purposes of detection, while the recovered correlation lengths should be read as resolution-limited, increasingly so along the sky minor axis as the inclination grows.

\subsection{Constructing the Null}
\label{sec:inclined:null}

As remarked upon in the previous subsection, judicious choice of null model can absorb many of the biases and systematics introduced by the analysis pipeline. The velocity residuals are measured relative to a background model of the disk rotation. For an inclined disk that model depends on the inclination, position angle, source center, systemic velocity and emission surface. Errors in any of these leave smooth, coherent residuals with low azimuthal wavenumbers that extend across the full annulus \citep[e.g.,][]{Yen_Gu_2020}. Tested against a noise-only null, such residuals would be correctly identified as excess structure, yet they are artifacts of the adopted geometry rather than properties of the velocity field. For a face-on disk the rotation has no line-of-sight component, so the analysis is largely insensitive to these parameters; for an inclined disk it is not. The null should therefore include the uncertainty in the background model, for example by drawing the geometric parameters from their posterior distributions and adding the resulting residuals to each noise realization.

\subsection{Velocity Projection}
\label{sec:inclined:projection}

For disks viewed almost face-on, the velocity perturbations observed will be dominated by the vertical motions. As one moves to more inclined sources, the projection of the in-plane motions, $\delta v_\phi$ and $\delta v_r$, will contribute more to the observed variance. However, the projection of the in-plane components will lead to azimuthally varying magnitudes, unlike for $\delta v_z$ which is azimuthally constant. Attributing observed variations in the velocity field to real local variations in velocity structure rather than changes in projection is a major hurdle in these analyses. To keep this discussion focused, we consider only fields that are statistically homogeneous in azimuth.

Under certain assumptions regarding the scale of the velocity structures, we can gain some insight. In the limit of large-scale structure, we see that $\delta v_\phi$ and $\delta v_r$ contributions reverse sign across the minor and major axes, respectively, imprinting an anti-correlation on the azimuthal structure function at separations beyond $\Delta\phi = 90\degr$. On smaller scales, we can expect that the projection is locally only a rescaling of the velocity strength. As such, we would expect that the short-lag heuristics $\hat{\ell}_r$ and $\hat{\ell}_s$ should be the least affected, while the large-lag behavior of the azimuthal structure function, and with it the anisotropy $\hat{\mathcal{A}}$, should be the most affected. Because the per-annulus $S_2$ averages over azimuth, recovering the modulation itself requires measuring $S_2$ in azimuthal sectors, which we leave to future work.

Localized perturbations with arbitrary velocity structure, such as those driven by embedded planets, are not described by such a statistical model. As noted in Section~\ref{sec:numerical_simulations}, they are better identified with feature-detection techniques \citep{Izquierdo_ea_2026} and, where they would otherwise dominate the structure function, masked before computing $S_2$.

As the result described in Section~\ref{sec:numerical_simulations} were based on disks viewed face-on, they only trace the vertical motions and thus will not be directly applicable to sources believed to host these instabilities but viewed at different inclinations.

\section{Conclusions}
\label{sec:conclusions}

We developed and validated a framework for detecting and characterizing correlated structure in protoplanetary disk velocity residuals using the second-order structure function. Working with correlated Gaussian random fields drawn natively on the disk polar grid (Section~\ref{sec:morphological_models}), we introduced a set of quick empirical heuristics alongside a complementary parametric fit (Section~\ref{sec:structure_function}), and built a pipeline that de-noises at the level of $S_2$ and assesses significance against noise-only null ensembles (Section~\ref{sec:noisy_models}). We then applied the complete pipeline to synthetic velocity maps of three (magneto-)hydrodynamical instabilities drawn from numerical simulations (Section~\ref{sec:numerical_simulations}), and to observations of TW~Hya (Section~\ref{sec:twhya}), and tested how the method extends to inclined disks (Section~\ref{sec:inclined}). This functionality has been included in the latest release of the \texttt{eddy} package. Our main conclusions are:
\begin{itemize}
    \item In the presence of noise, excess variance above the noise model is detected robustly against the matched noise-only null across the full range of noise conditions considered, from $z \approx 7$ where the noise equals the field in amplitude and has twice its correlation length to $z \approx 120$ where it is a quarter of the amplitude on half the scale.
    \item Characterizing that variance is a more challenging prospect when the intrinsic variations are similar in magnitude to noise. The azimuthal length $\hat{\ell}_s$ is the most readily detected property, and only where the field is genuinely anisotropic. The anisotropy $\hat{\mathcal{A}}$ is only marginally recovered, and only for the zero-pitch anisotropic model, while $\hat{\ell}_r$, $\hat{\alpha}_r$ and $\hat{\alpha}_{\phi}$ are consistent with the null models.
    \item We find that parameter recovery is parameter-dependent. The radial correlation length is recovered most reliably, while the azimuthal length is the more biased of the two rather than the less precise. Its fractional scatter matches that of the radial length, but its median fractional bias is roughly twice as large, and it is biased low for the pitched model. The amplitude $\sigma$ can be recovered without systematic bias, its precision limited by the sampling variance. A fitted radial slope $\alpha_r$ is a reliable indicator of genuine radial scaling, but not a sensitive test of stationarity. Across the four parameters the fit is less biased than the empirical heuristics more often than not, and decisively so for the correlation lengths.
    \item Applied to velocity maps from one numerical simulation each of the gravitational (GI), magneto-rotational (MRI) and vertical-shear (VSI) instabilities, the pipeline detects excess structure decisively in all three ($z = 266$, $40$ and $27$ for the GI, MRI and VSI), and distinguishes these three simulations by their structural heuristics. The observationally-degraded GI spiral and VSI rings are the hardest pair to tell apart ($4.1\sigma$, and the pair most sensitive to the noise model). Whether the mechanisms themselves are separable in general, rather than these particular realizations, remains to be tested on a suite of simulations spanning the variation within each class.
    \item The parametric Gaussian description is adequate only where the underlying structure is close to monotonic (here, the MRI). Coherent features such as the GI spiral arm or the quasi-periodic VSI rings drive the structure function above its saturating plateau, which no Gaussian kernel can reproduce. In this regime the empirical heuristics remain the more reliable descriptor, and a dedicated coherent-ridge model is required to characterize the feature itself.
    \item Separately from that misspecification, the parametric model struggles even where it does describe the field. With the pitch freed, none of the five GRF shape parameters is informative for these simulations: all fall within the noise-only null, leaving the amplitude $\sigma$ as the only well-constrained parameter. The pitch is the clearest case. Its signature vanishes as the field approaches isotropy, so that it is recovered only for $\ell_\phi / \ell_r \gtrsim 4$, and the kernel is exactly invariant under $(\ell_r, \ell_\phi, \beta) \rightarrow (\ell_\phi, \ell_r, \beta \pm 90\degr)$. A free-pitch fit therefore constrains the correlation ellipse but not the labels attached to its axes, and the empirical heuristics, measured on-axis, carry no such ambiguity.
    \item An application to $^{12}$CO observations of TW~Hya decisively reveals correlated velocity structure. The structure is elongated azimuthally but only marginally resolved radially. The $m=1$ component carries roughly half of the variance above the noise, yet azimuthally coherent perturbations persist once it is removed. The example also shows that the choice of background model directly changes the amplitude and azimuthal heuristics, so must be reported alongside them, and that the null must be forward-modeled, as the formal uncertainties on the velocity map overestimate its noise.
    \item For inclined disks, a null drawn through the same deprojection as the data absorbs the inclined geometry: structure remains robustly detected to $i = 60\degr$, and the heuristics are recovered within a few per cent of their face-on values at $30\degr$, whereas at $60\degr$ the stretched beam smooths the field, lowering $\hat\sigma$, lengthening $\hat\ell_r$ and compressing $\hat{\mathcal{A}}$. Beyond geometry, the null for an inclined source must also carry the uncertainty in the background rotation model, and the projection of in-plane motions is expected to modify the large-lag azimuthal structure, and with it $\hat{\mathcal{A}}$, most strongly.
\end{itemize}

Throughout this work we have treated the noise as correlated, beam-convolved white noise -- the simplest description adequate for a methods paper. Real interferometric noise is more structured, and the application to TW~Hya (Section~\ref{sec:twhya}) shows that the null must be forward-modeled from the data rather than taken from the formal uncertainties on the velocity map. A complete treatment, propagating the noise through imaging and the construction of the velocity map, is left to future work. Two departures are worth naming. The autocorrelation of interferometric image noise is the point spread function rather than its autocorrelation, so a beam-convolved white-noise null is correlated on a scale $\sqrt{2}$ too long; and the point spread function is not Gaussian, carrying side lobes that imprint anti-correlation no analytic Gaussian form can reproduce. Neither affects the results of Sections~\ref{sec:structure_function}--\ref{sec:numerical_simulations}, which are measured against matched nulls drawn from the same model as the injected noise, but both matter when the null is built for a real image. For TW~Hya we adopt the $\theta_b / \sqrt{2}$ correlation scale, while the side lobes remain untreated. As real image noise is correlated on a shorter scale than the beam-convolved noise of Section~\ref{sec:noisy_models}, the detection statistics presented there are likely conservative.

\begin{acknowledgments}
We thank the anonymous referee for a thorough reading of the manuscript and for providing helpful comments. We thank James Owen for a reading of the manuscript and thoughtful suggestions. This paper makes use of the following ALMA data: ADS/JAO.ALMA\#2013.1.00387.S and ADS/JAO.ALMA\#2018.A.00021.S. ALMA is a partnership of ESO (representing its member states), NSF (USA) and NINS (Japan), together with NRC (Canada), NSTC and ASIAA (Taiwan), and KASI (Republic of Korea), in cooperation with the Republic of Chile. The Joint ALMA Observatory is operated by ESO, AUI/NRAO and NAOJ. The National Radio Astronomy Observatory is a facility of the U.S. National Science Foundation operated under cooperative agreement by Associated Universities, Inc.
This work was supported by NASA's Exoplanets Research Program through grant No. 80NSSC26K1149.
MB-A acknowledges funding from the European Union's Horizon Europe research and innovation programme under the Marie Skłodowska-Curie grant agreement No.~101281707 (InnerDisk). Views and opinions expressed are however those of the author(s) only and do not necessarily reflect those of the European Union or the European Research Executive Agency. Neither the European Union nor the granting authority can be held responsible for them.
\end{acknowledgments}

\appendix

\section{Per-heuristic significance}
\label{sec:appendix_significance}

\begin{figure*}
    \centering
    \includegraphics[width=\textwidth]{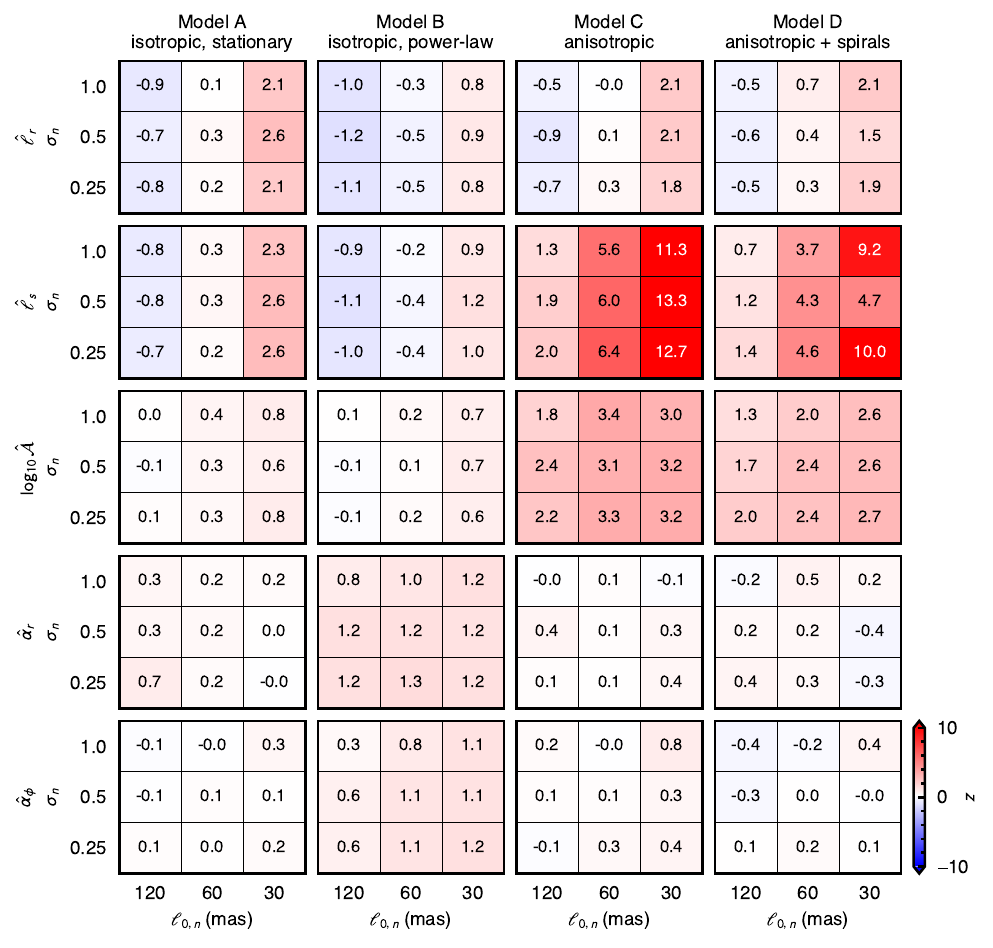}
    \caption{Signed significance $z$ of each structural heuristic (rows) for every velocity model (columns), measured against the matched noise-only null. Each panel is the same $3\times3$ noise grid as Figure~\ref{fig:significance_maxz} -- noise amplitude $\sigma_n$ increasing upward, spatial scale $\ell_{0,n}$ decreasing to the right. The color scale is linear in $z$ and saturates at $|z| = 10$; the dashed marks on the color bar are the $|z| = 3$ significance level, and near-white cells within them are consistent with noise. This is the per-heuristic decomposition of Figure~\ref{fig:significance_maxz}, which plots the largest $|z|$ across these five rows in each cell. The amplitude $\hat\sigma$ is excluded, as in Figure~\ref{fig:significance_maxz}; its significance is the subject of Figure~\ref{fig:noise_significance}.}
    \label{fig:significance_grid}
\end{figure*}

Figure~\ref{fig:significance_maxz} reports, for each model and noise condition, only the single structural heuristic that departs most strongly from the matched noise-only null. Figure~\ref{fig:significance_grid} gives the full breakdown behind that collapse: the signed $z$-score of each of the five structural heuristics individually, on the same models and noise grid, so that the significance figure can be read heuristic by heuristic. It makes explicit that the detections in the anisotropic models (C and D) are driven entirely by the azimuthal correlation length $\hat{\ell}_s$, which reaches $z = 13.3$ at the smallest noise scale, and that the only other heuristic to contribute at all is the anisotropy: $\log_{10}\hat{\mathcal{A}}$ reaches $|z| = 3.1$--$3.4$ in five Model C cells, while $\hat{\alpha}_r$ and $\hat{\alpha}_\phi$ stay below $|z| = 3$ in every cell of the grid and Models A and B are consistent with the null throughout.

\software{\texttt{astropy} \citep{astropy_2013, astropy_2018, astropy_2022},
          \texttt{bettermoments} \citep{Teague_Foreman-Mackey_2018},
          \texttt{eddy} \citep{eddy},
          \texttt{matplotlib} \citep{Hunter_2007},
          \texttt{numpy} \citep{Harris_ea_2020},
          \texttt{scipy} \citep{Virtanen_ea_2020}}

\bibliography{main}{}
\bibliographystyle{aasjournalv7}

\end{document}